\documentclass[journal=jacsat,manuscript=article]{achemso}

\usepackage{chemformula} % Formula subscripts using \ch{}
\usepackage[T1]{fontenc} % Use modern font encodings

\usepackage{graphicx}% Include figure files
\usepackage{dcolumn}% Align table columns on decimal point
\usepackage{bm}% bold math

\usepackage{amsmath}
\usepackage{amssymb}
\usepackage{color}
\usepackage[normalem]{ulem}
\usepackage{hyperref}
\usepackage{comment}
\author{Fabio Leoni}
\affiliation{Department of Physics, Sapienza University of Rome, P.le Aldo Moro 5, 00185 Rome, Italy}
\author{Carles Calero}%
\affiliation{Secci\'o de
 F\'isica Estad\'istica i Interdisciplin\`aria--Departament de F\'{i}sica de la Mat\`{e}ria Condensada, Universitat de Barcelona, 
 \& Institut de Nanoci\`{e}ncia i Nanotecnologia (IN2UB), Universitat de Barcelona, C. Mart\'{i} i Franqu\`{e}s 1, 08028 Barcelona, Spain}%
\author{Giancarlo Franzese}%
\affiliation{Secci\'o de
 F\'isica Estad\'istica i Interdisciplin\`aria--Departament de F\'{i}sica de la Mat\`{e}ria Condensada, Universitat de Barcelona, 
 \& Institut de Nanoci\`{e}ncia i Nanotecnologia (IN2UB), Universitat de Barcelona, C. Mart\'{i} i Franqu\`{e}s 1, 08028 Barcelona, Spain}%
\email{fabio.leoni@uniroma1.it.}

\title{Nanoconfined fluids: Uniqueness of water compared to other liquids}


\begin{document}

\begin{abstract}
Nanoconfinement can drastically change the behavior of liquids, puzzling us with counterintuitive properties. Moreover, it is relevant in applications, including decontamination and crystallization control. However,  it still lacks a systematic analysis for fluids with different bulk properties. Here we fill this gap. We compare, by molecular dynamics simulations, three different liquids in a graphene slit pore: (A) A simple fluid, such as argon, described by a Lennard-Jones potential; (B) An anomalous fluid, such as a liquid metal, modeled with an isotropic core-softened potential; (C) Water, the prototypical anomalous liquid, with directional hydrogen bonds. We study how the slit-pore width affects the structure, thermodynamics, and dynamics of the fluids.  We check that all the liquids, as expected, have a) free-energy minima---hence mechanical stability---for widths that are optimal to accommodate fluid layers, b) mechanically-unstable free-energy maxima for intermediate widths, c) an effective wall-wall repulsion at sub-optimal widths, i.e., for under-sized slit-pores, d) a fluid-mediated attraction for over-sized slit-pores, e) oscillations in diffusion, proportional to those in free-energy, between slower (at the free-energy minima) and faster diffusion (at the free-energy maxima). However, the nature of the free-energy minima for the three fluids is quite different. In particular, i) only for the simple liquid all the minima are energy-driven, while the structural order in the minima increases with decreasing slit-pore width; ii) only for the isotropic core-softened potential all the minima are entropy-driven, while the energy in the minima increases with decreasing slit-pore width; iii) only the water has a changing nature of the minima: the monolayer minimum is entropy-driven, at variance with the simple liquid, while the bilayer minimum is energy-driven, at variance with the other anomalous liquid. Also, water diffusion has a large increase for sub-nm slit-pores, becoming faster than bulk. Instead, the other two fluids have diffusion oscillations much smaller than water.  Both the isotropic liquids slow down for decreasing slit-pore width, with the simple liquid near freezing at sub-nm confinement. Our results clarify that nanoconfined water is unique compared to other (simple or anomalous) fluids under similar confinement, and are possibly relevant in nanopores applications, e.g., in water purification from contaminants. 
\end{abstract}

%%%%%%%%%%%%%%%%%%%%%%%%%%%%%%%%%%%%%%%%%%%%%%%%%%%%%%%%%%%
%
%     INTRODUCTION
%
%%%%%%%%%%%%%%%%%%%%%%%%%%%%%%%%%%%%%%%%%%%%%%%%%%%%%%%%%%%
 
\section{\label{sec:introduction}Introduction}
Fluids under nanoconfinement are challenging for understanding because they can show
properties that are quite different compared to their bulk counterpart
\cite{schoen1998, Truskett01, Mittal2008, De-Virgiliis2008,
  Rzysko2010,  schnell2011,  schnell2012, Paul2012, Stewart2012,
  Krott2013, Karan2012}.  
For example, they form layers parallel to the confining
surfaces \cite{bhushan1995}, and, when the confinement width is 
ultrathin, the layers can solidify in peculiar structures \cite{jones1972}.
In the case of nanoconfined water, freezing can happen both above \cite{Zangi2003, algara-siller2015, acs.jpcc.8b09840} or below \cite{doi:10.1021/nn901554h,doi:10.1021/acsnano.0c03161} the bulk melting temperature depending on the confining system. Simulations of a monoatomic water model nanoconfined to form only two layers show even dynamical oscillations between the liquid phase and ice \cite{doi:10.1021/acsnano.8b03403}.
Nanoconfined fluids are relevant for their implications in life science and nanotechnology \cite{Bellissent95,
  Mashl2003, Mallamace2008, cicero2008, Giovambattista09,
  castrillon2009, castrillon2009b, Mancinelli2009, Gallo2010,
  Rzysko2010, Han2010,  delosSantos2011, giovambattista2012,
  Nair2012, Ferguson2012,Schiro2009, Biedermann2013, Franzese2013,
 doi:10.1021/acsnano.0c02984}, and for applications such as
purification of fluids forced through microporous carbon materials \cite{Abraham:2017vk, Zhou:2018aa, Hirunpinyopas:2020wj}
%is key in industrial processes and applications, e.g.,  medical uses
nano-lubrication \cite{D0NR08121C},
or isotope separation in nuclear power technology
%, where diffusion
%properties are sensitive to the pore width and geometry
\cite{C3CP44414G}.
The fabrication of nanoscale membranes \cite{geim2013} allow to
investigate transport properties at the molecular level, revealing
fast permeation of water through carbon nanotubes 
\cite{hummer2001,majumder2005,holt2006} and through graphene oxide
 membranes \cite{joshi2014}, which can be used for filtration of
 complex mixtures and water disinfection and desalination.  

Confined fluids have been studied extensively by numerical simulations
in various geometries, including slit, tubular, and cubic pores, with
flat or rough walls or with different wall permeabilities, finding relationships between pore size
and selectivity \cite{doi:10.1063/1.464356}. 
In particular, computer simulations show  that
nano-confinement may influence the dynamical properties
of fluids. For example,  liquid films of nonpolar
molecules, confined  between two solid walls, undergo an
abrupt change in the diffusion constant and support
shear  \cite{thompson1992,diestler1993}, or freeze to a solid as the structured wall \cite{rhykerd1987}, when the confinement reduces to a few molecular layers.
Experiments confirm the liquid-to-solid transition for
simple organic solvents (cyclohexane, octamethylcyclotetrasiloxane, and toluene) under confinement 
when decreasing from seven to six molecular
layers  \cite{klein1995,Klein:1998aa}.

Molecular dynamics (MD) simulations of a Lennard-Jones (LJ) liquid in
slit pores, with widths from 2 to 12 molecular diameters
and structureless walls, show a weak
increase of the local parallel diffusion  for the particles initially
within the first layer near the wall \cite{Winkler:1996aa}.
By varying pore width at constant chemical
potential,
both  parallel diffusion coefficient and solvation
force oscillate and saturate to the bulk value for  widths greater than
10 molecular diameters \cite{doi:10.1063/1.449375, PhysRevE.98.052115}.
Moreover, both for a LJ liquid  \cite{Zhang:2011aa} and a LJ binary equimolar mixture
\cite{Hannaoui:2013aa}, the self-diffusion coefficient reduces when the confining scale decreases
or the interaction of the fluid with the walls increases.
However, 
these results are at variance with those for  simple gases confined in  carbon nanotubes, where the diffusion coefficient is larger for smaller nanotube diameters \cite{Mao:2000aa, Mao:2001aa, PhysRevLett.89.185901, Bhatia:2005aa}.

Also, for 
anomalous liquids \cite{Bordin2012} and water  in  carbon nanotubes, the diffusion coefficient changes in a  non-monotonic way and the flow can be enhanced \cite{Thomas:2008bh, Thomas2009, Qin2011}
with  decreasing nanotube diameters \cite{Allen1999}, especially for diameters below 1 nm \cite{Mashl2003, Ye2011, Barati-Farimani2011, Zheng2012}.
 On the other hand, previous simulations of water confined in nanotube of different diameters show that the diffusion along the axes decreases for smaller diameters \cite{Mart:2002aa}.
 
 Contraddictory results have been found also for
the shear viscosity of water confined in a graphene nonotube. It monotonically increases for increasing channel diameter \cite{hongfei2011,doi:10.1063/1.3592532} or oscillates and decreases for increasing slit pore width, depending on the specific water model \cite{doi:10.1021/acsnano.6b00187}.

It is, therefore, worth asking how these varieties of different results depend on the details of the fluid interactions or the confining geometry.
For example, Striolo finds a relevant difference between the diffusion of simple fluids and water in molecular sieves \cite{Striolo:2006fv}.
While the first is dominated by concerted events in which multiple molecules move simultaneously due to the spatial mismatches between pore-fluid and fluid-fluid attractive interactions, the ballistic diffusion of water clusters is a consequence of long-lasting hydrogen bonds (HBs) \cite{Striolo:2006fv}.

Here, we deepen this question and ask which property of water nanoconfined in a graphene-like slit pore is unique and which is shared with other anomalous liquids, or even normal liquids.
To this end, we perform  MD simulations of the LJ fluid and the Continuous Shouldered Well (CSW) anomalous fluid \cite{Fr07a,OFNB08,vilaseca:084507,leoni2014,leoni2016} under slit-pore confinement. The  CSW fluid is a coarse-grained model for fluids, including
liquid metals or complex liquids \cite{Vilaseca2011}, with water-like properties associated to the presence of  two length scales \cite{OFNB08,vilaseca:084507}, such as the hydrophobic effect
\cite{Hus:2013kx}. 
It is used, also, to model 
hydroxyl groups interactions in methanol \cite{Hus:2014aa, Hus2014, Desgranges:2018aa}, and water-hydroxyl groups interactions in water/methanol mixtures \cite{Marques:2020aa}.
A potential similar to the CSW has been used to study the effect of macromolecular crowders in biological media with high concentration of proteins, polysaccharides or nucleic acids \cite{Blanco:2018aa,D0SM01475C}.
Yet, the CSW fluid has not a water-like entropy behavior, as all the other two-length scales isotropic potential, because it has no directional interactions \cite{Vilaseca2011,russo2021}. 
We compare the behavior of these two liquids with that of TIP4P/2005 water,
in which, instead, the specific geometry of four charges induce the electrostatic interactions responsible for the HBs along preferred directions.

The TIP4P/2005 water in a graphene slit-pore has free-energy extrema determining diffusion oscillations, with free-energy/diffusion minima  for wall-wall distances fitting complete layers, down to one, and maxima at intermediate distances \cite{calero2020}. In particular, the  free-energy minimum for a monolayer originates from an increase of water disorder, despite the corresponding water internal energy increases. For the bilayer, instead, the free-energy minimum is dominated by a minimum in internal energy per water molecule with a larger order  \cite{calero2020}. The latter, with a full HB network, is the minimum with the largest mechanical stability \cite{calero2020}, rising the question if it would be so also in a fluid without HBs.

%%%%%%%%%%%%%%%%%%%%%%%%%%%%%%%%%%%%%%%%%%%%%%%%%%%%%%%%%%%
%
%     RESULTS
%
%%%%%%%%%%%%%%%%%%%%%%%%%%%%%%%%%%%%%%%%%%%%%%%%%%%%%%%%%%%

\section{Results and discussion}
\label{sec:results}

\begin{figure}%[t!]
(a) \includegraphics[clip=true,width=6cm]{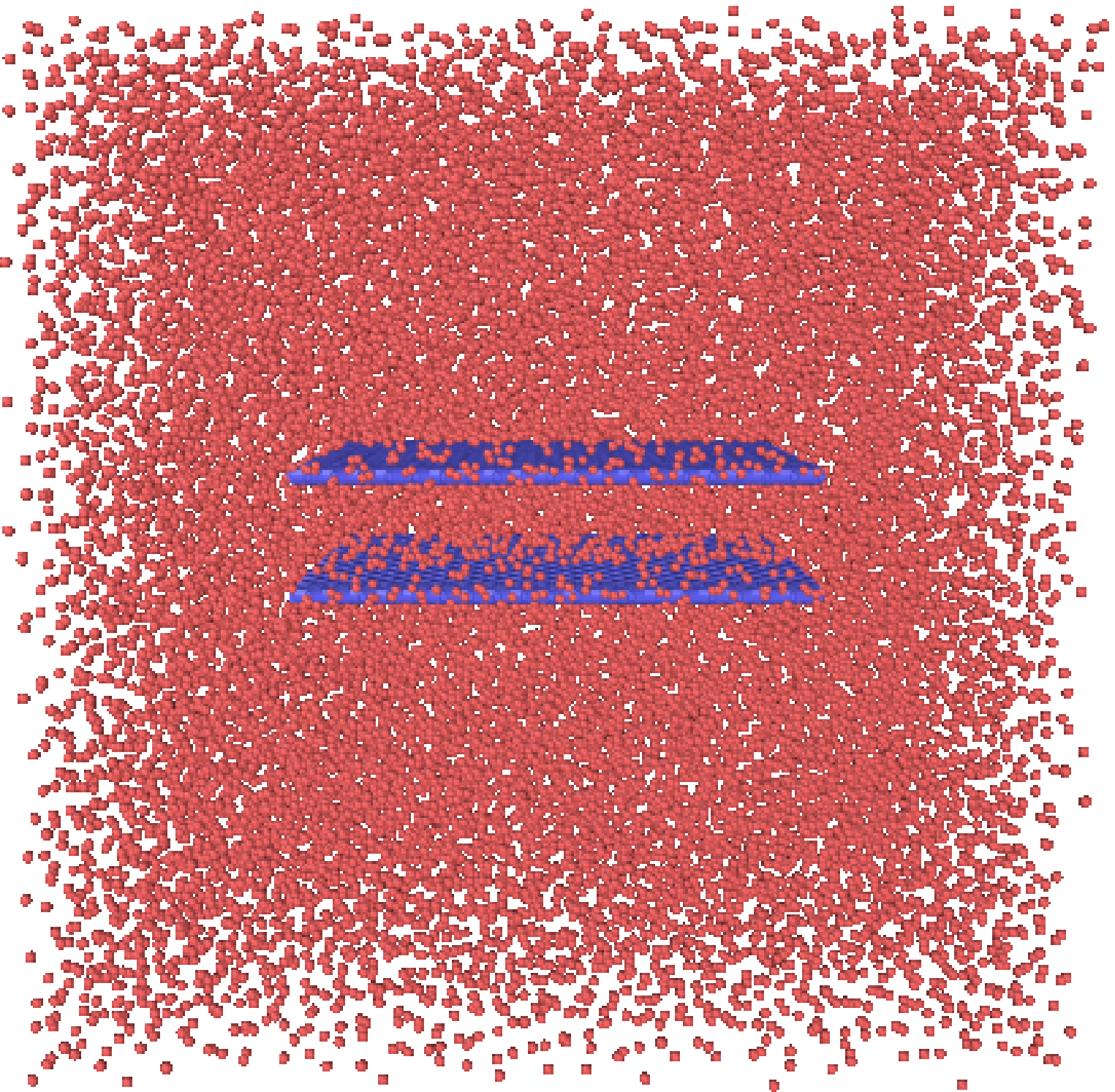}
\hspace{1cm}
(b) \includegraphics[clip=true,width=6cm]{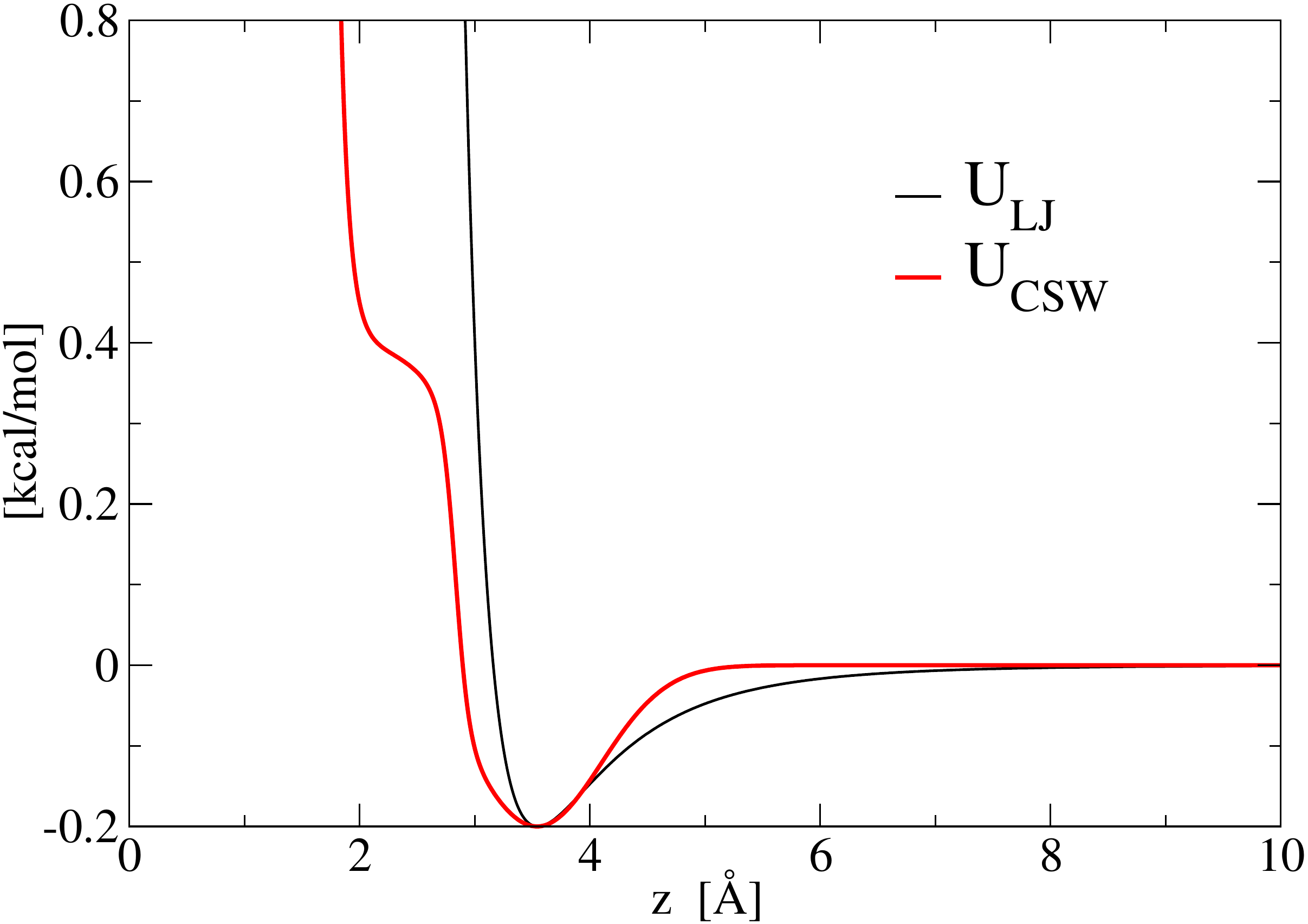}
\caption{Simulation geometry and isotropic potentials.
(a) 
Snapshot of the simulation box with 
  $N_{\rm tot}=25000$ CSW particles at $\rho=0.036$~{\AA}$^{-3}$ and $T=100$ K, 
  with a graphene slit-pore with width  
  $\delta=11$~{\AA} and area $A\simeq  25$ nm$^2$.
(b) 
LJ (black line) and CSW (red line) inter-particle potentials for systems A and B, respectively, as described in the text.}  
%CSW rho0.2 T80 d11}
\label{fig:snapshot}
\end{figure}

We perform MD simulations (see Methods section for details) of three different fluids surrounding a  graphene slit-pore (Fig.~\ref{fig:snapshot}): 
(A) A simple fluid, described by a LJ potential; 
(B) An anomalous fluid, with water-like properties but different from water, modeled with the isotropic CSW potential; 
(C) TIP4P/2005 water. 
For each fluid, we simulate a box, with periodic boundary conditions, with 
a slit-pore, centered at the origin of the reference system, made of two parallel graphene sheets of fixed area  $A$, separated by a distance $\delta$ and positioned a $z_{p_{\pm}}=\pm \delta/2$.
We consider nanoscopic slit-pores of width ranging from  
$\delta=6.5$~{\AA} to $\delta=17$~{\AA}, with $0.5$~{\AA} increments.
To reduce the edge effects of the walls, we compute the properties only of the confined fluids with coordinates 
$-L_x^s/2<x<L_x^s/2$,
$-L_y^s/2<y<L_y^s/2$ and 
$z_{p_-}<z<z_{p_+}$, i.e.,  
within a central sub-volume $V^s\equiv A^s \times \delta$, where $A^s\equiv L_x^sL_y^s$, with $L_x^s=L_y^s=30$~{\AA} for the isotropic potentials (A) and (B) and 15~{\AA} for the TIP4P/2005 water (C).

\subsection{Structure}

We first analyze how the confinement affects the structure of the isotropic liquids.
As other confined liquids, the LJ and CSW fluids form layers parallel to the walls, displaying peaks in  their density profiles $\rho_z(z)$ along the normal direction $z$  (Figs.~S1, S2).   
% (Fig.\ref{fig:Rhoz}).   
The number of layers increases with the distance $6.5$ \AA$ \leq \delta\leq 17$ \AA\  between the
plates, going from 1 to 4 for the LJ , and from 1 to 5 for the CSW.
The presence of two characteristic length scales in the CSW potential leads to the formation of complex patterns \cite{leoni2014} and structured peaks, especially at higher densities (not shown), that are absent in the LJ. 
%Fig.~\ref{fig:Rhoz}). 

\begin{figure*}[t!]
\begin{center}
\begin{minipage}{7.8cm}
\includegraphics[clip=true,width=7.5cm]{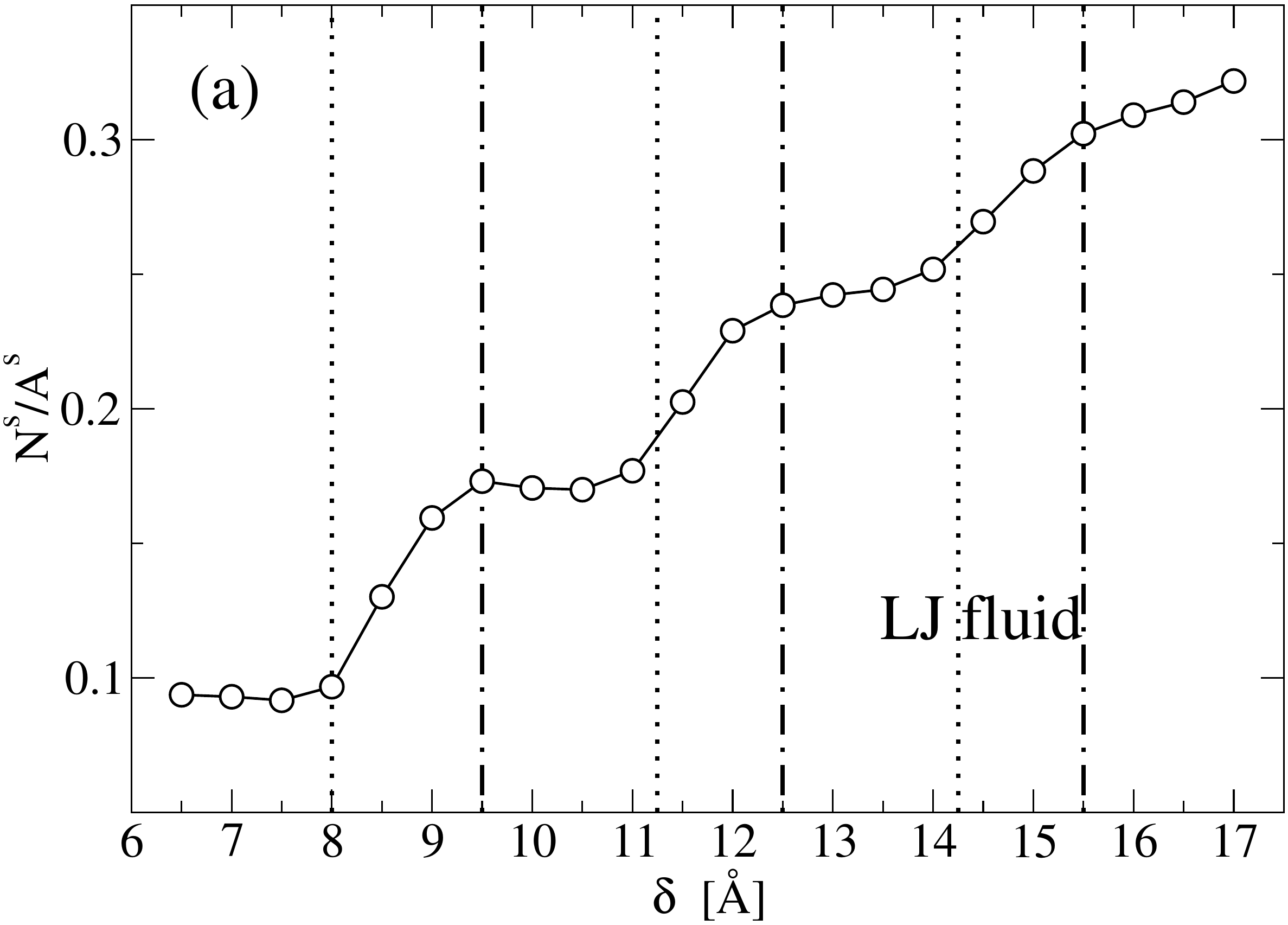}\hspace{-0.1cm} 
\includegraphics[clip=true,width=7.8cm]{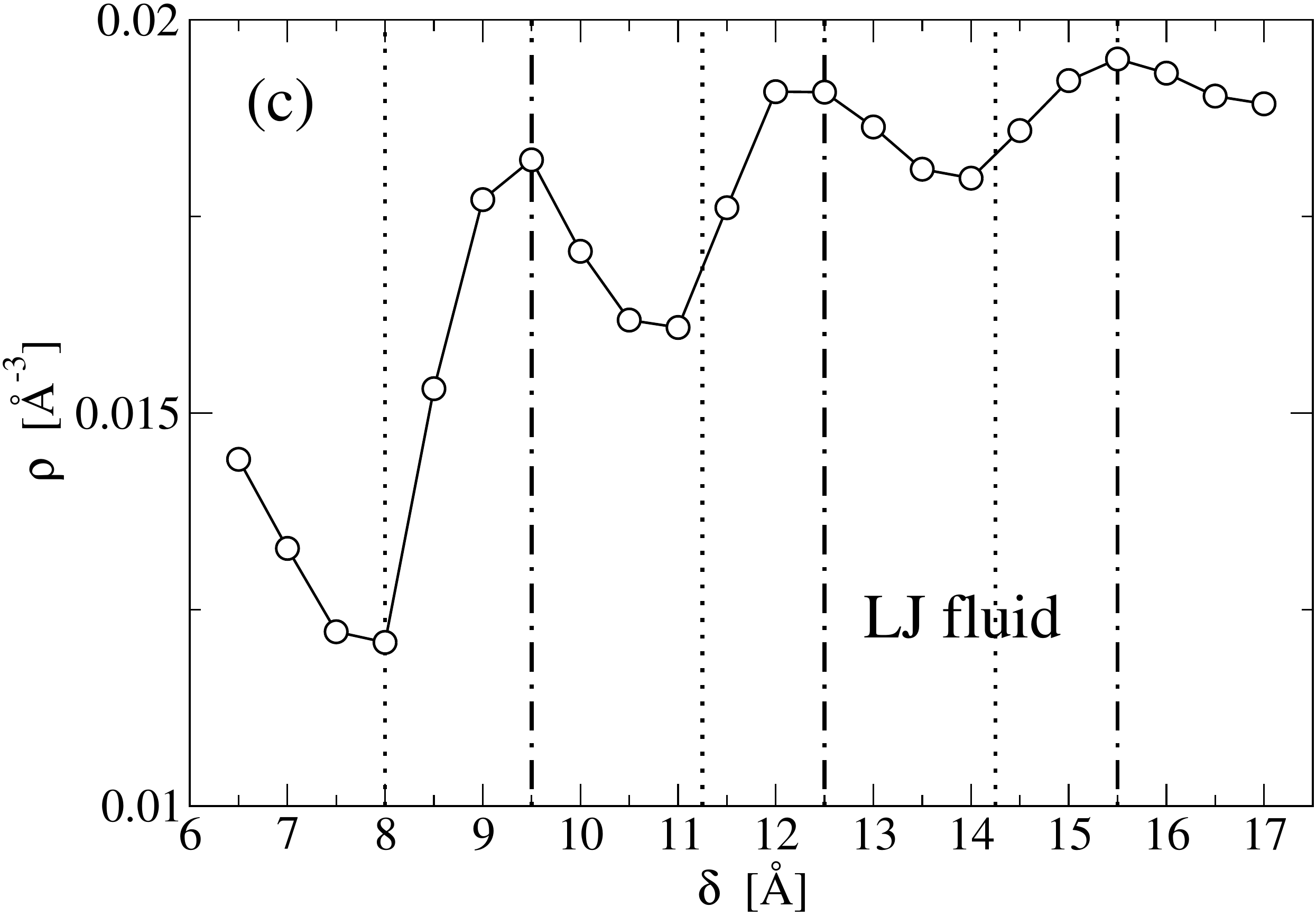} 
\end{minipage}
\begin{minipage}{7.8cm}
\includegraphics[clip=true,width=7.5cm]{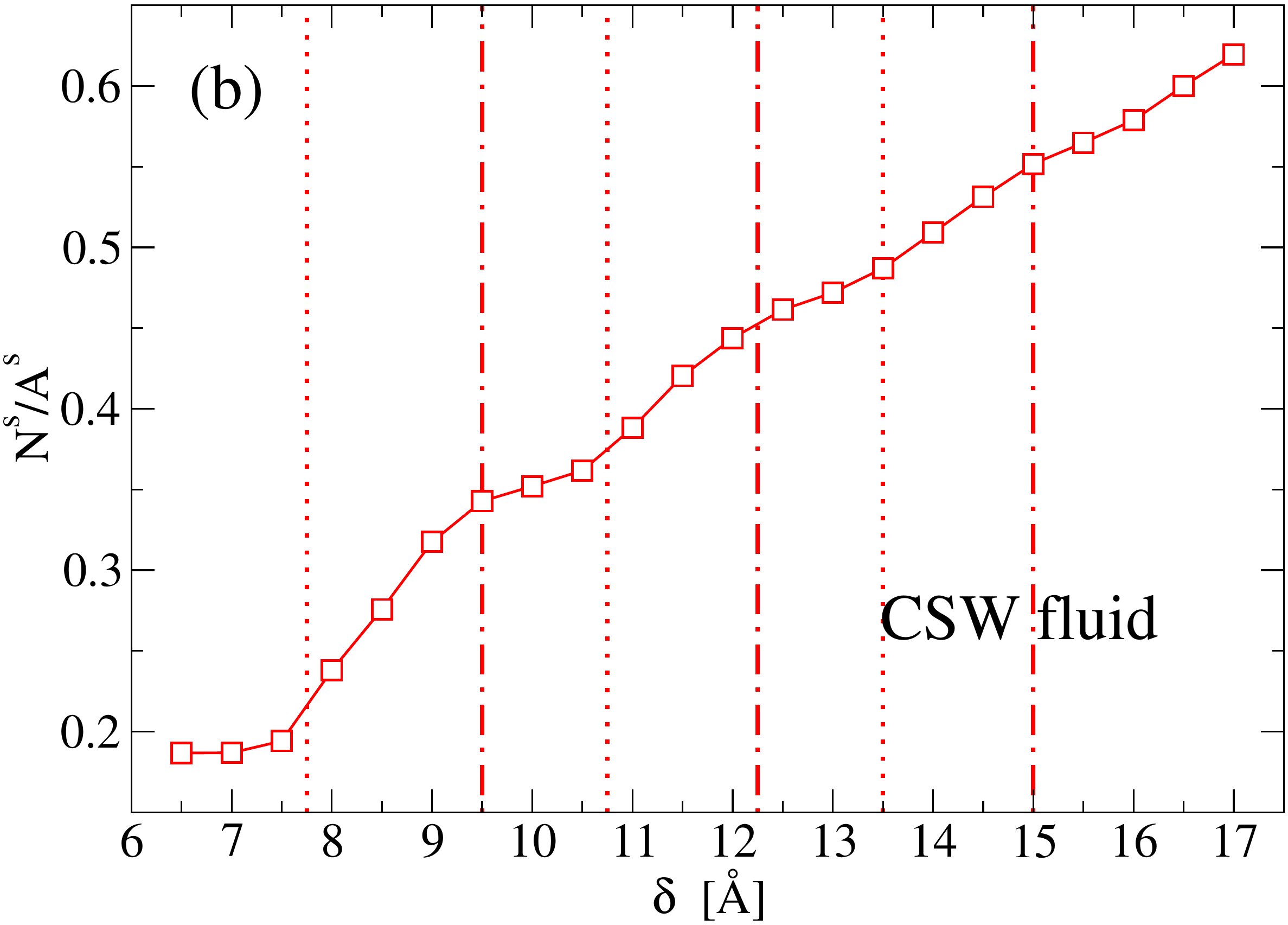} 
\includegraphics[clip=true,width=7.7cm]{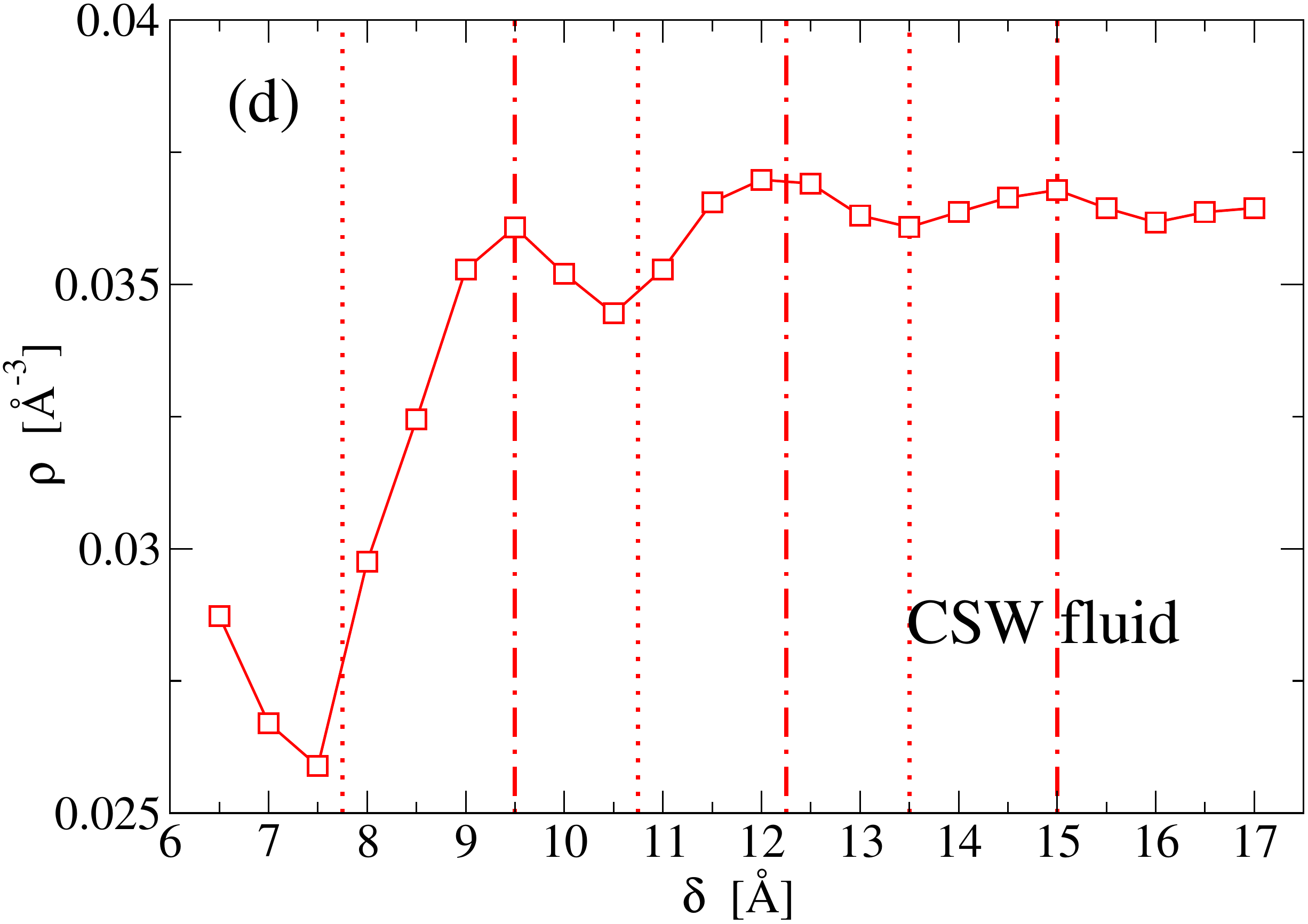} 
\end{minipage}
\end{center}
\caption{Slit-pore  acceptance capacity, $N^s/A^s$, and mean density $\rho$ for the confined isotropic fluids  as a function of the plate separation $\delta$.  
(a, b) The LJ parameters are: $\rho_{LJ}=0.023$~\AA$^{-3}$ and $T_{LJ}=100K$.
(c, d) The CSW parameters are: $\rho_{CSW}=0.036$~\AA$^{-3}$
%($\rho^*_{CSW}=0.2$) 
and $T_{CSW}=100K$. 
%($T_{CSW}^*=1.0$).}
 The (dotted and dot-dashed) vertical lines are defined in Fig.\ref{fig:diff} and mark approximately the extrema (maxima and minima, respectively) of the fluid diffusion coefficient $D_\parallel$ in layers parallel to the plates.
}     
\label{fig:Nconf}
\end{figure*}

We find that the slit-pore {\it acceptance capacity}, defined as the number of confined particles $N^s/A^s$
normalized by the sub-volume area $A^s$,
for both  fluids  has a steplike behavior as  a function of $\delta$ (Fig.\ref{fig:Nconf} a, c).
These steps resemble what has been found for water under similar confinement \cite{Engstler:2018ab, calero2020} and it is a result of the layering. 
% FIG: Nconf, Nconf/d 
Indeed, the comparison with Figs. S1 and S2 shows that a step starts at values of $\delta$ where a new layer appears (e.g., for the LJ: $\delta\simeq 8$~{\AA}, 11.25 ~{\AA}, 14.25 ~{\AA}; for the CSW: $\delta\simeq 7.75$~{\AA}, 10.75 ~{\AA}, 13.5 ~{\AA}). The steps smoothen for larger $\delta$  as the  confined fluid becomes less structured.

We can emphasize this behavior by analyzing how the mean density $\rho$ of the  fluid within the pore changes with $\delta$ (Fig.\ref{fig:Nconf} c, d). It shows oscillations, approaching the bulk value for increasing $\delta$.
The mean density reaches minima at those separations where, for an increase of $\delta$, 
the fluid starts a new layer and the particles are sucked inside the pore from the reservoir.

For intermediate separations both liquids fill the layers up to reach maxima in $\rho$, corresponding to optimal plates distances (e.g., for the LJ: $\delta\simeq 9.5$~{\AA}, 12.5 ~{\AA}, 15.5 ~{\AA}; for the CSW: $\delta\simeq 9.5$~{\AA}, 12.25 ~{\AA}, 15 ~{\AA}) where the density profiles $\rho_z(z)$ display well-formed peaks, sharper and higher than those for slightly different  $\delta$ (Figs. S1, S2).
A further increase of $\delta$, up to the value for a new layer, does not increase the acceptance capacity (plateaus in Fig. \ref{fig:Nconf} a, b), leading to a new minimum in $\rho$. 
For the CSW fluid, the plateaus of the acceptance capacity and the oscillation in $\rho$ 
are less pronounced and shifted towards smaller values of $\delta$ (Fig.\ref{fig:Nconf} b, d) as a consequence of the interaction soft-core.

As we will discuss in the following sections, these steps and oscillations are associated to oscillatory behaviors in dynamics (vertical lines in Fig.\ref{fig:Nconf}), hydration forces and thermodynamics. In particular,  
the relation between structure and entropy can be emphasized by calculating 
the translational order parameter \cite{To00,Tr00} in each layer $i$, defined as
%the direction parallel to the confining plates associated to each layer $i$. $t_{\|}^i$ accounts for the spatial order in layer $i$, defined as
\begin{equation}
t_{\|}^i\equiv\int_0^{\infty}|g_{\|}^i(\xi^i)-1|d\xi^i
\end{equation}
where 
$\xi^i\equiv r_{\|}(\rho^i_\|)^{1/3}$ is the longitudinal distance $r_{\|}$, parallel to the walls,
in units of the mean interparticle separation $(\rho^i_\|)^{-1/3}$,
$\rho^i_\|$ is the fluid density  in the layer $i$,
% projected on the plane parallel to the walls, 
 and $g^i_{\|}(\xi^i)$ is
the longitudinal radial distribution function. For an ideal gas,
$g_{\|}(\xi)=1$, hence there is no translational order ($t_{\|}=0$). 

% Order Parameter t for LJ
\begin{figure*}[t!]
\includegraphics[clip=true,width=7.5cm]{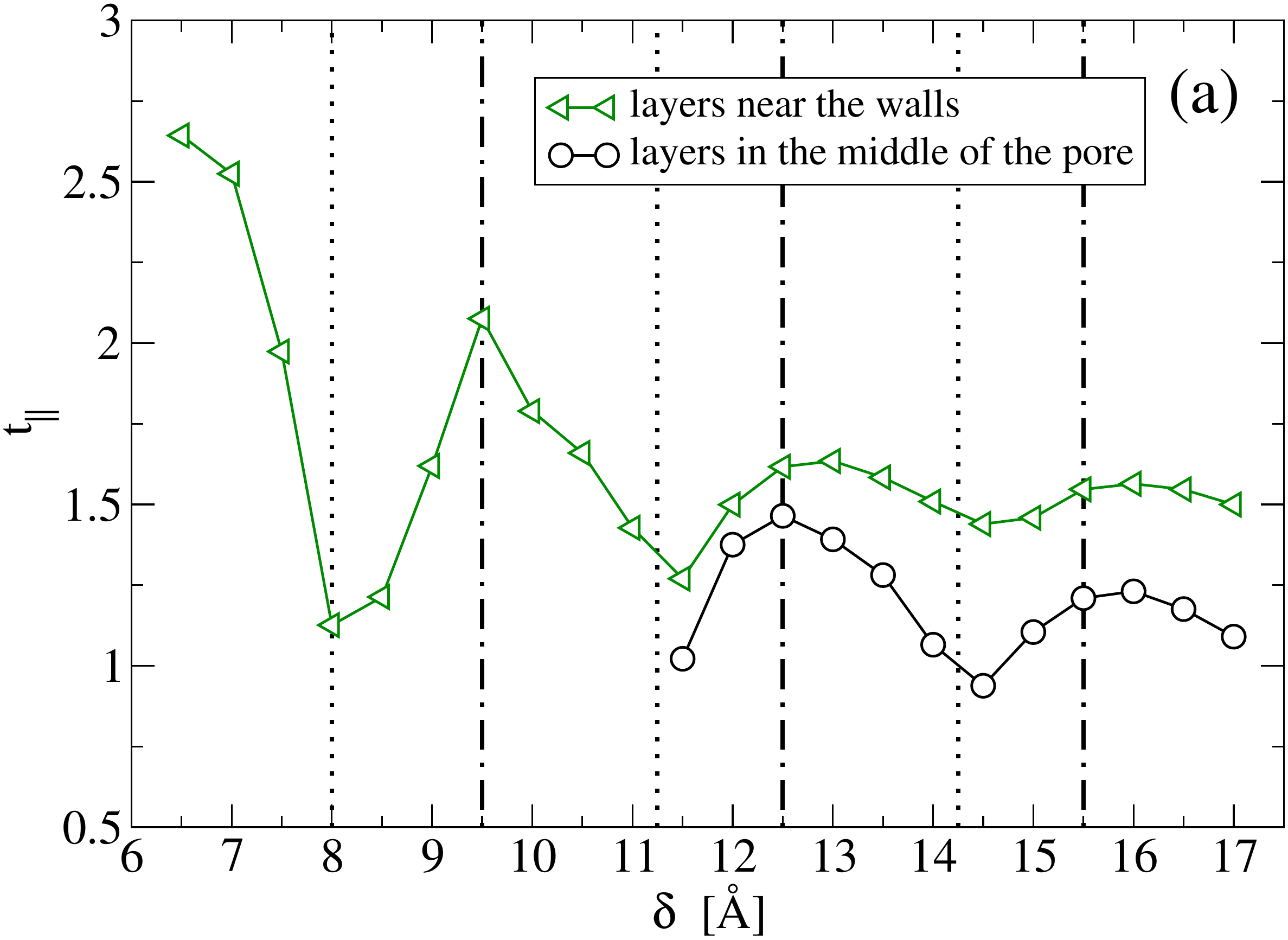}
\hspace{0.2cm}
\includegraphics[clip=true,width=7.5cm]{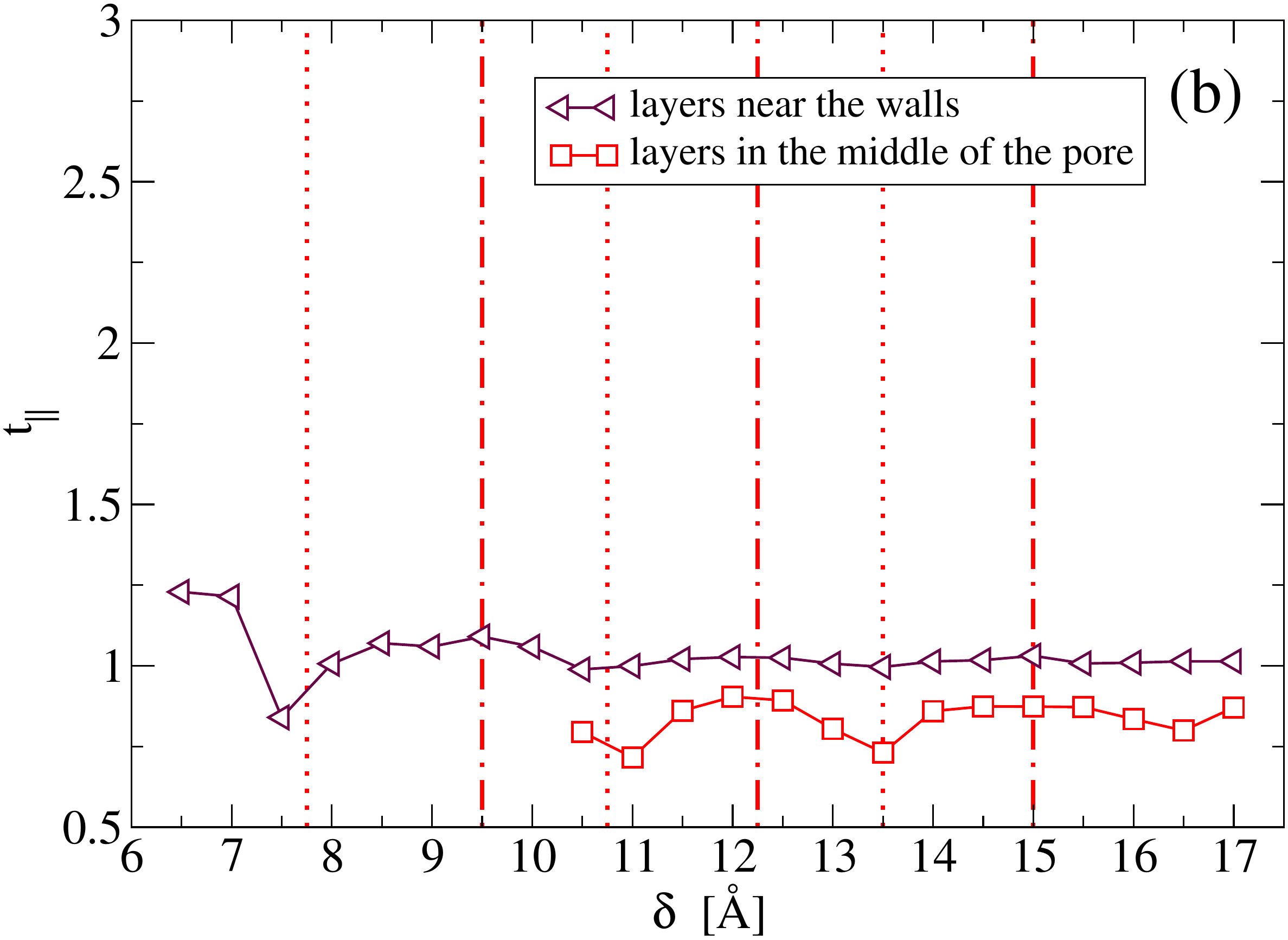}
\caption{Longitudinal translational order parameter $t_{\|}$ for each layer of confined isotropic fluids as a 
function of the plate separation $\delta$. The parameter is calculated for the layers in contact with the walls (triangles) and, separately, for the other layers (circles for LJ, squares for CSW). 
The thermodynamic conditions for the LJ (a) and the CSW (b) and the vertical lines are the same as in Fig.~\ref{fig:Nconf}.
}
% The fluid parameters are: $\rho_{LJ}=0.023~\AA^{-3}$ and $T_{LJ}=100K$
%for the LJ fluid; $\rho_{CSW}=0.036~\AA^{-3}$ ($\rho^*_{CSW}=0.2$) and
%$T_{CSW}=100K$ ($T_{CSW}^*=1.0$) for the CSW fluid.}   
\label{fig:torderp}
\end{figure*}

We calculate the parameter separately for the layers in contact with the walls and for the other layers (Fig.~\ref{fig:torderp}), finding that both oscillate with $\delta$ and that the contact layers are always more ordered than the inner layers.
However, they have maxima and minima at the same values of $\delta$, showing that the plate separation can regulate the order in the whole confined fluid.

In particular, the layers are more ordered when the mean density $\rho$ of the confined fluid is maximum, i.e., when the  $\delta$ is optimal for well-formed layers.
The fluid order decreases when the mean density $\rho$ is minimum, corresponding to the appearance/disappearance of a  new layer.

For small $\delta$, when the slit-pore contains only one or two layers of the fluid, $t_{\|}^i$ has larger oscillations, although for the CSW liquid the variation is weaker, as a consequence of its soft-core. In general, 
the CSW is always less ordered than  the LJ at the same plate separation $\delta$. However, for both isotropic fluids, the structural oscillations, due to the layering, are determining the translational order, hence the entropy, of the confined liquid and are correlated to its dynamics (vertical lines in Fig.~\ref{fig:torderp}). In the next section we show how we locate the vertical lines marking the extrema in the dynamics.

\subsection{Dynamics}

\begin{figure*}%[t!]
\includegraphics[clip=true,width=7.5cm]{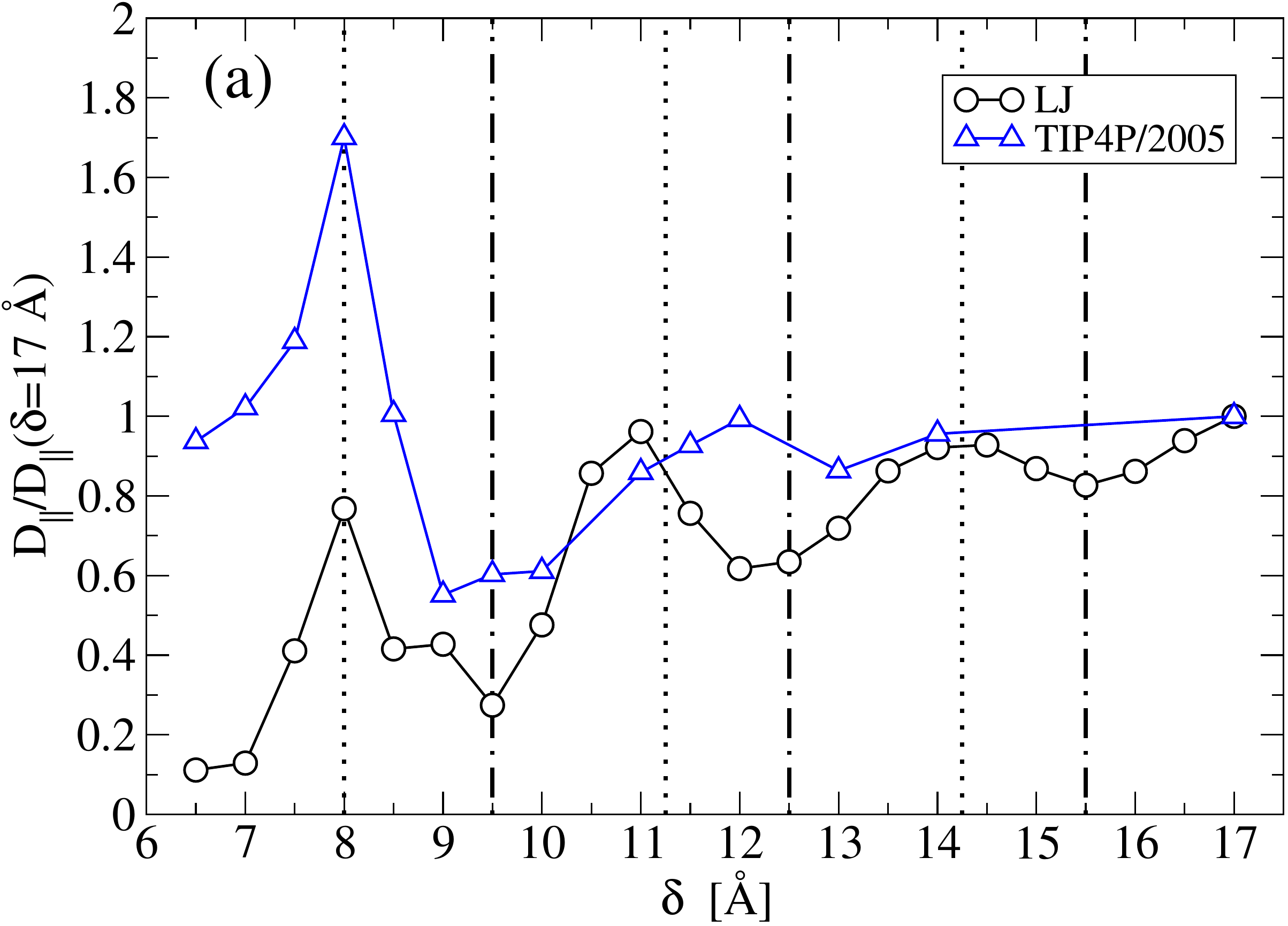}
\hspace{0.2cm}
\includegraphics[clip=true,width=7.5cm]{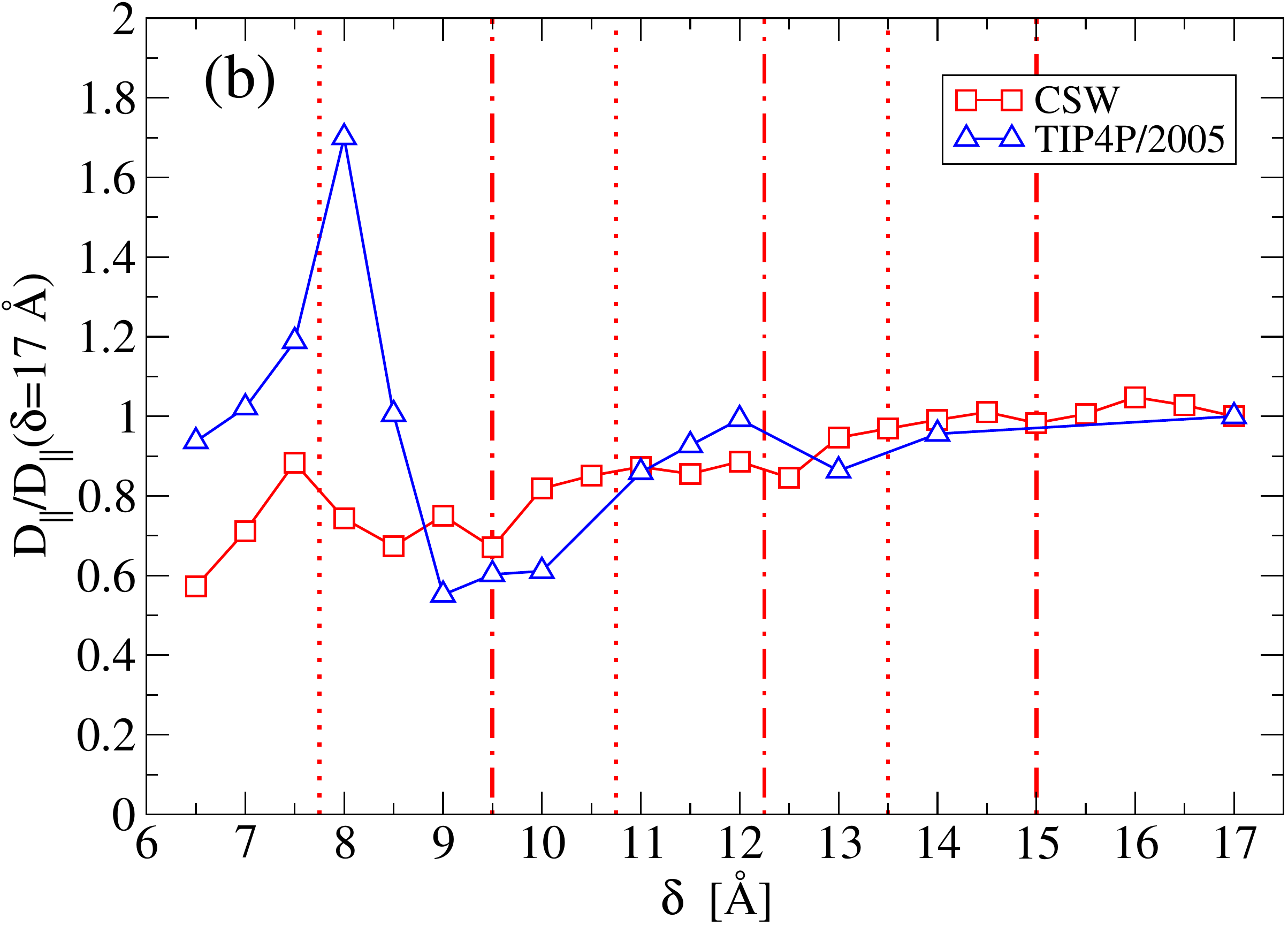}
\caption{Longitudinal diffusion coefficient $D_{\|}$, normalized to its large-$\delta$ value, for the three fluids in a slit-pore,  as a function of the plate separation $\delta$.  Comparison of the TIP4P/2005-water (blue triangles) \cite{calero2020} with (a) the LJ (black circles), and (b) the CSW (red squares). 
In both panels vertical lines mark, approximately, maxima (dotted lines) and minima (dot-dashed lines) for the isotropic fluid (see text). The value of $D_{\|}$
at $\delta=17$~\AA\ is $\simeq 23$ nm$^2$/ns for both the LJ and the CSW, and is $\simeq 1.9$ nm$^2$/ns for the TIP4P/2005-water \cite{calero2020}.}
\label{fig:diff}
\end{figure*}

Next, we analyze how the confinement affects the thermal motion, in the direction parallel to the plates, by calculating the longitudinal diffusion coefficient, 
$D_{\|}$,  for our three prototypical liquids, as a function of the plates inter-distance $\delta$,
 with 
\begin{equation}
D_{\|} \equiv \lim_{\tau\rightarrow\infty}\langle(\Delta 
r_{\|}(\tau))^2\rangle/(4
\tau),
\end{equation}
where 
\begin{equation}\label{equ:MSD}
\langle (\Delta r_{\|}( t-t_0))^2\rangle
\equiv 
\langle (r_{\|}(t-t_0)-r_{\|}(t_0))^2\rangle
\end{equation}
is the longitudinal mean square displacement, 
with $r_{\|}\equiv (x^2+y^2)^{1/2}$, 
$\tau\equiv t-t_0$ is the time spent in the confined sub-region $V^s$ 
by a particle entering  $V^s$ at $t_0$,
and
$\langle\cdots \rangle$ is the average over 10 time-intervals, each made of $10^4$ MD steps (Fig.\ref{fig:diff}).

% Parallel diffusion

We observe that the three fluids share three properties:
\begin{itemize}
\item[(i)]  $D_{\|}$ is not monotonic as a function of $\delta$, 
\item[(ii)] $D_{\|}$-oscillations are larger for smaller $\delta$, and 
\item[(iii)] $D_{\|}$ is not monotonic as a function of the average density $\rho$ (Fig.~S4 in Supplementary Material), indicating anomalous behavior.
\end{itemize}
Because only the water \cite{Agarwal2011} and the CSW fluid \cite{OFNB08} can show anomalous diffusion in the bulk, while the LJ fluid cannot, we conclude that these three properties are not necessarily related to the bulk anomalies.  

The property (iii) resembles recent results for other confined anomalous-fluid models where it was attributed to the competition of two interaction length-scales \cite{Krott2013,Krott:2013ux}, the appearance of amorphous phases \cite{Krott:2013ux} or the reentrance of the melting line \cite{bordin2018}. However, here we find it also for the simple fluid without competing length-scales,  amorphous phases or reentrant melting, showing that the presence of confinement is enough to get the property (iii), as well as the (i) and (ii), in the three fluids.

Nevertheless, there are relevant differences among the three cases.
\begin{itemize}
\item[(iv)] For both isotropic (LJ and CSW)  fluids, $D_{\|}$ oscillates but it is always smaller than its bulk value, with a decreasing trend for decreasing $\delta$. In the case of water, instead, the fastest diffusion is reached at $\delta\approx 8$~\AA, between one and two confined layers \cite{calero2020}.
\item[(v)]  Although both the isotropic fluids have, for the selected state-point, a diffusion coefficient $D_{\|}\simeq$ 23 nm$^2$/ns for $\delta=17$~\AA\ ($\simeq 10$ times larger than the value for water), 
the oscillations of  $D_{\|}$ in the three fluids are quite different:  $\approx 90$\% for LJ, $\approx 40$\% for CSW, and $\approx 70$\% for water. 
\item[(vi)] For sub-nm confinement ($\delta<10$~\AA),  the  minima and maxima of the $D_{\|}$-oscillations are approximately located at the same separations for the three liquids. However, the oscillations mismatch for $\delta>10$~\AA, being opposite at $\delta\approx 12.5$~\AA\, especially comparing LJ and water.
\end{itemize}

The sub-nm correspondence is better between LJ and water because the size of the LJ particles is equal to that of the LJ-component of the water model, while the CSW soft-core reduces the effective size of the particles and smoothens out the effect.
The matching of the oscillations for $\delta<10$~\AA\ confirms \cite{calero2020} that the steric hindrance (layering)  has a major role in determining the dynamics under confinement of a simple liquid, as well as an anomalous liquid.
However, this mechanism is no longer enough to rationalize the behavior for larger confinement, as emphasized by the mismatch for $\delta>10$~\AA\ and the differences highlighted in (iv) and (v). This observation calls for an alternative explanation for the peculiar dynamics of confined water. As we will show in the following, it is related to the unique properties of the water HBs.

\subsection{Thermodynamics}

\subsubsection{Hydration Pressure}

\begin{figure*}[t!]
\includegraphics[clip=true,width=7.5cm]{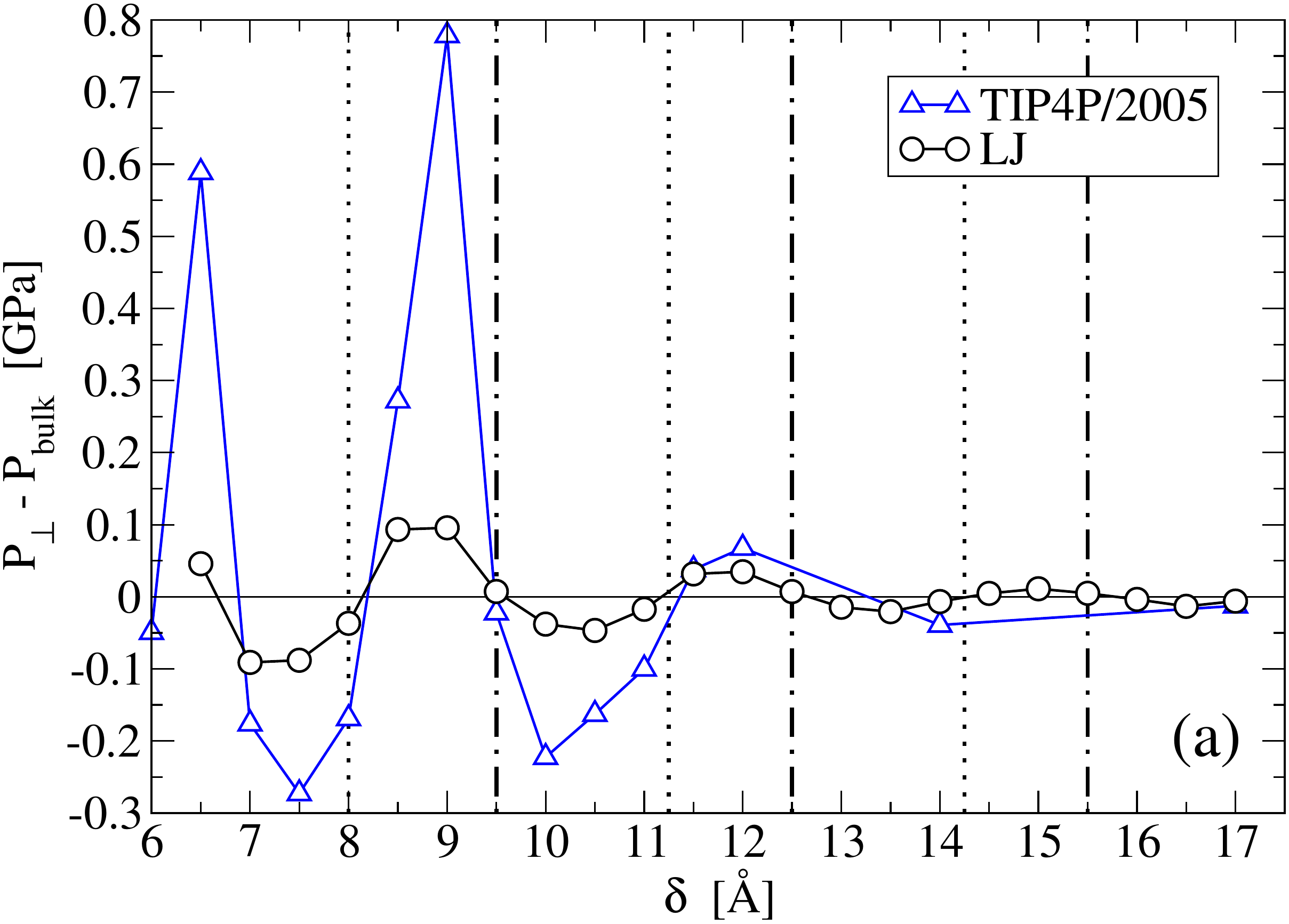}
\hspace{0.2cm}
\includegraphics[clip=true,width=7.5cm]{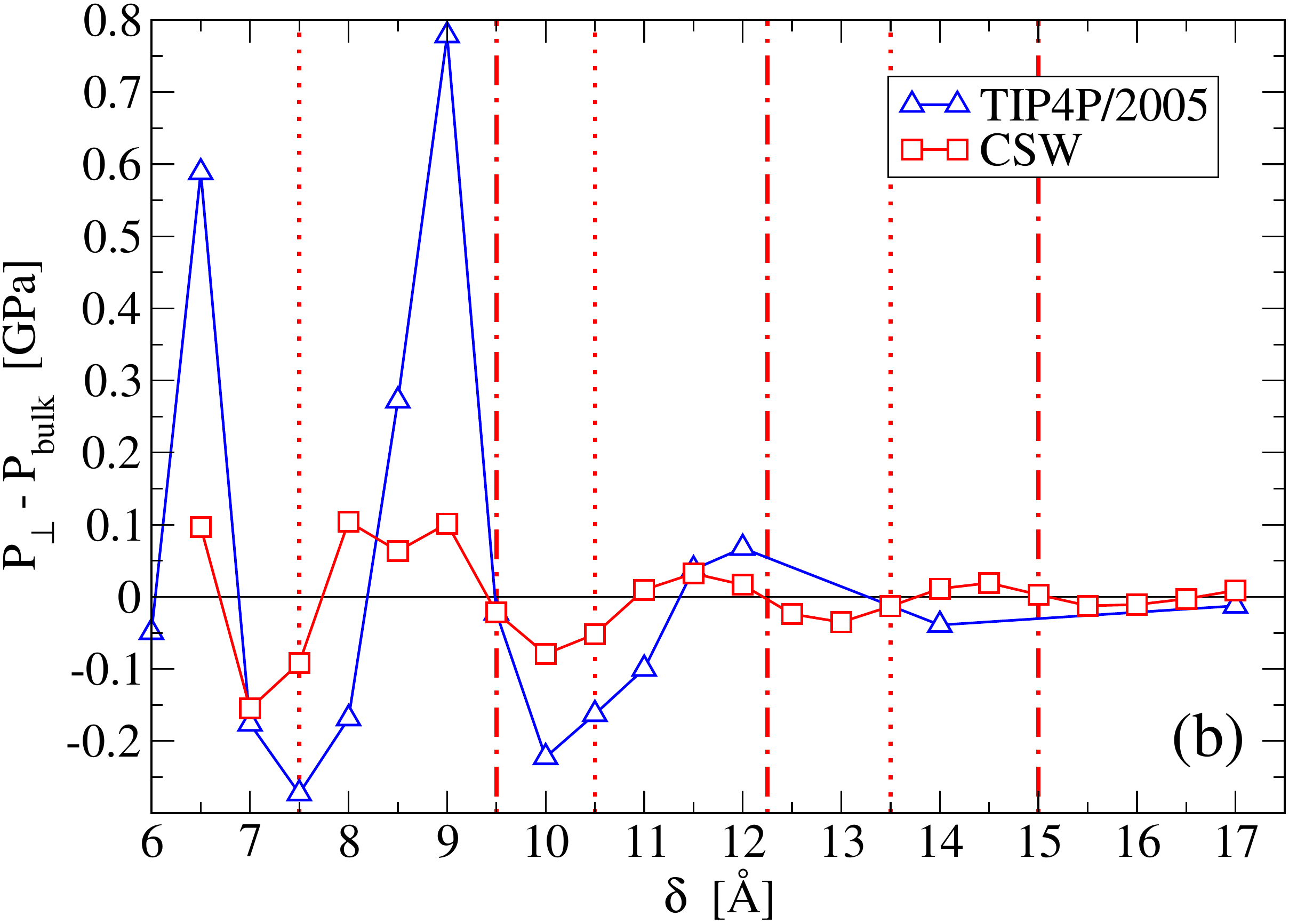}
\caption{Hydration pressure, $P_{\rm hydr}$, for the confined fluids as a function of the plate separation $\delta$. Comparison of the TIP4P/2005-water (blue triangles) \cite{calero2020} with (a) the LJ (black circles), and (b) the CSW (red squares). In both panels the (dotted and dot-dashed) vertical lines are those indicated in Fig.\ref{fig:diff}, marking approximately, the extrema (maxima and minima, respectively) of $D_\parallel$ for the isotropic fluids. We observe that all the lines in panels (a) and (b) approximately cross the zeros of  $P_{\rm hydr}$ for the LJ (a) and the CSW fluids (b), respectively
%The fluid parameters are: $\rho_{LJ}=0.023~\AA^{-3}$ and $T_{LJ}=100K$ for the LJ fluid; $\rho_{CSW}=0.036~\AA^{-3}$ ($\rho^*_{CSW}=0.2$) and
%$T_{CSW}=100K$ ($T_{CSW}^*=1.0$) for the CSW fluid;
%$\rho_{TIP4P/2005}=0.033$~{\AA}$^{-3}$ and $T=300K$ for the TIP4P/2005.
}
\label{fig:pressure}
\end{figure*}

To better understand the differences between the three confined fluids,
we calculate the hydration pressure, $P_{\rm hydr} \equiv P_{\perp}-P_{\rm bulk}$, as a function of $\delta$.
Here, $P_{\perp}$ is the normal pressure that the confined fluid exerts on the plates (Fig.\ref{fig:pressure}).

Over the entire range of $\delta$ explored here, we find that $P_{\rm hydr}$ oscillates for the three fluids and approaches zero at $\delta=17$~{\AA}. Hence, at large plates separation the confined fluids behave as in the bulk.
When  $P_{\rm hydr}>0$ there is an effective repulsion between the plates, while when $P_{\rm hydr}<0$ there is a fluid-mediated attraction.  In both cases, the walls are kept fixed in their position by our simulation constraints. The constraint is not necessary  when, instead, the walls are at equilibrium, with  $P_{\rm hydr}=0$. 

We observe that, for LJ and CSW fluids, the equilibrium $\delta$-values, at which $P_{\rm hydr}=0$, coincide, within our numerical precision, with the extrema of $D_{\|}$. 
Hence, the system is in equilibrium, not only when the thermal diffusion is minimal, but also when it is maximal. This suggests that the two equilibrium positions have a very different origin, as already observed in the case of water \cite{calero2020}. 

In particular, if $\delta^{\rm{MAX}~D_{\|}}_1$ and $\delta^{\rm{min}~D_{\|}}_1$ are the shortest distances for a maximum and a minimum $D_{\|}$, respectively, 
%(marked by vertical lines in all the figures of results as a function of $\delta$), 
displacing
 the pore-size from $\delta^{\rm{min}~D_{\|}}_1$ induces a change in pressure that tends to restore the wall-to-wall distance. Hence, $\delta^{\rm{min}~D_{\|}}_1$ corresponds to a distance of stable mechanical equilibrium. The opposite occurs around $\delta^{\rm{MAX}~D_{\|}}_1$, hence it corresponds to a distance of unstable mechanical equilibrium.

By decreasing $\delta$ from $\delta^{\rm{min}~D_{\|}}_1$ to $\delta^{\rm{MAX}~D_{\|}}_1$,  $P_{\rm hydr}$ increases  up to a maximum and, at intermediate distances, decreases  toward  $P_{\rm hydr}=0$.
 Hence, squeezing the fluid toward $\delta^{\rm{MAX}~D_{\|}}_1$ implies  a speedup of the thermal diffusion 
 and  a work against the effective wall-wall repulsion. 

At $\delta^{\rm{MAX}~D_{\|}}_1$ the fluid has maximum diffusion at unstable mechanical equilibrium, $P_{\rm hydr}=0$. 
%However, the equilibrium is unstable because 
Any further squeezing induces an attraction, $P_{\rm hydr}<0$, between the walls. 
In this case, the work to reduce $\delta$ is done by the fluid-mediated wall-wall attraction and slows down the thermal parallel diffusion.

Between the two equilibrium values $\delta^{\rm{MAX}~D_{\|}}_1$ and $\delta^{\rm{min}~D_{\|}}_1$, $P_{\rm hydr}$ for LJ and TIP4P/2005 liquids displays  a  single  peak, while the CSW fluid has  two close peaks.
This difference can be understood as a signature of the two competing length-scales of the CSW potential. Similar considerations hold for all $\delta^{\rm{MAX}~D_{\|}}_i$ and $\delta^{\rm{min}~D_{\|}}_i$, although we find only simple maxima of $P_{\rm hydr}$ for the CSW.
%%

%Comparing the $P_{\rm hydr}$ for the water with the two isotropic fluids, 
We observe that the water $P_{\rm hydr}$ has its largest maximum (repulsion) around $\delta\simeq 9$~\AA, corresponding to a confined bilayer, with a smaller maximum for the monolayer at $\delta\simeq 6.5$~\AA\ and the trilayer at $\delta\simeq 12$~\AA.
 For the isotropic fluids, instead, the maxima in $P_{\rm hydr}$  for the bilayer and the monolayer are approximately equal and larger than those for more layers, at least within our resolution. This observation suggests that the work to approach the walls for a water bilayer is larger than for a monolayer, while is it approximately the same for the isotropic fluids. This is consistent with the result showing that the water bilayer is more stable than the monolayer  \cite{calero2020}, and suggests that it is not for the isotropic fluids. To deepen this understanding, we calculate, and compare, the free energy of the confined fluids in the next section.

%, Energy and Entropy difference
\begin{figure*}[t!]
\begin{center}
\begin{minipage}{7.5cm}
\includegraphics[clip=true,width=7.5cm]{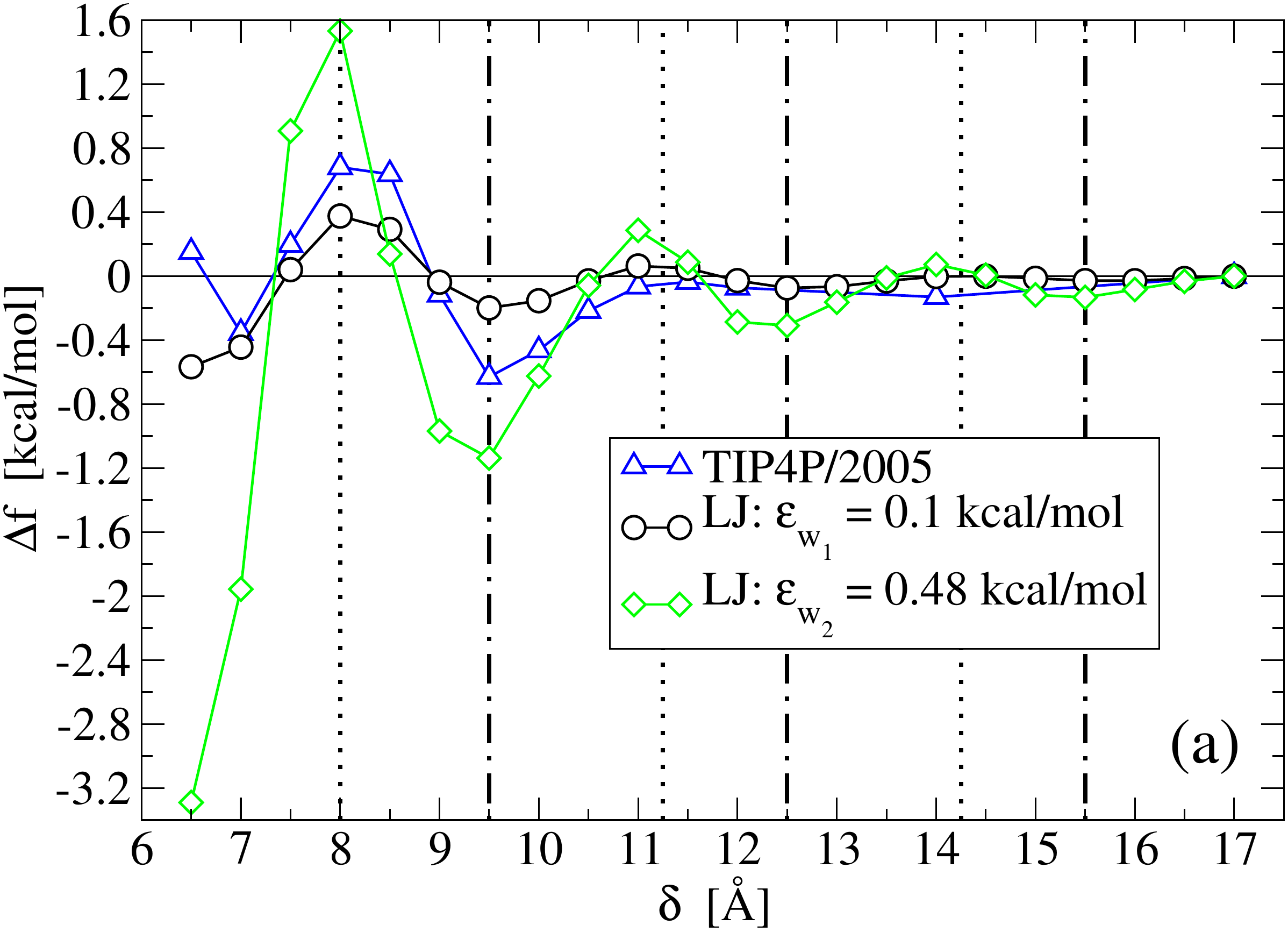}
\includegraphics[clip=true,width=7.5cm]{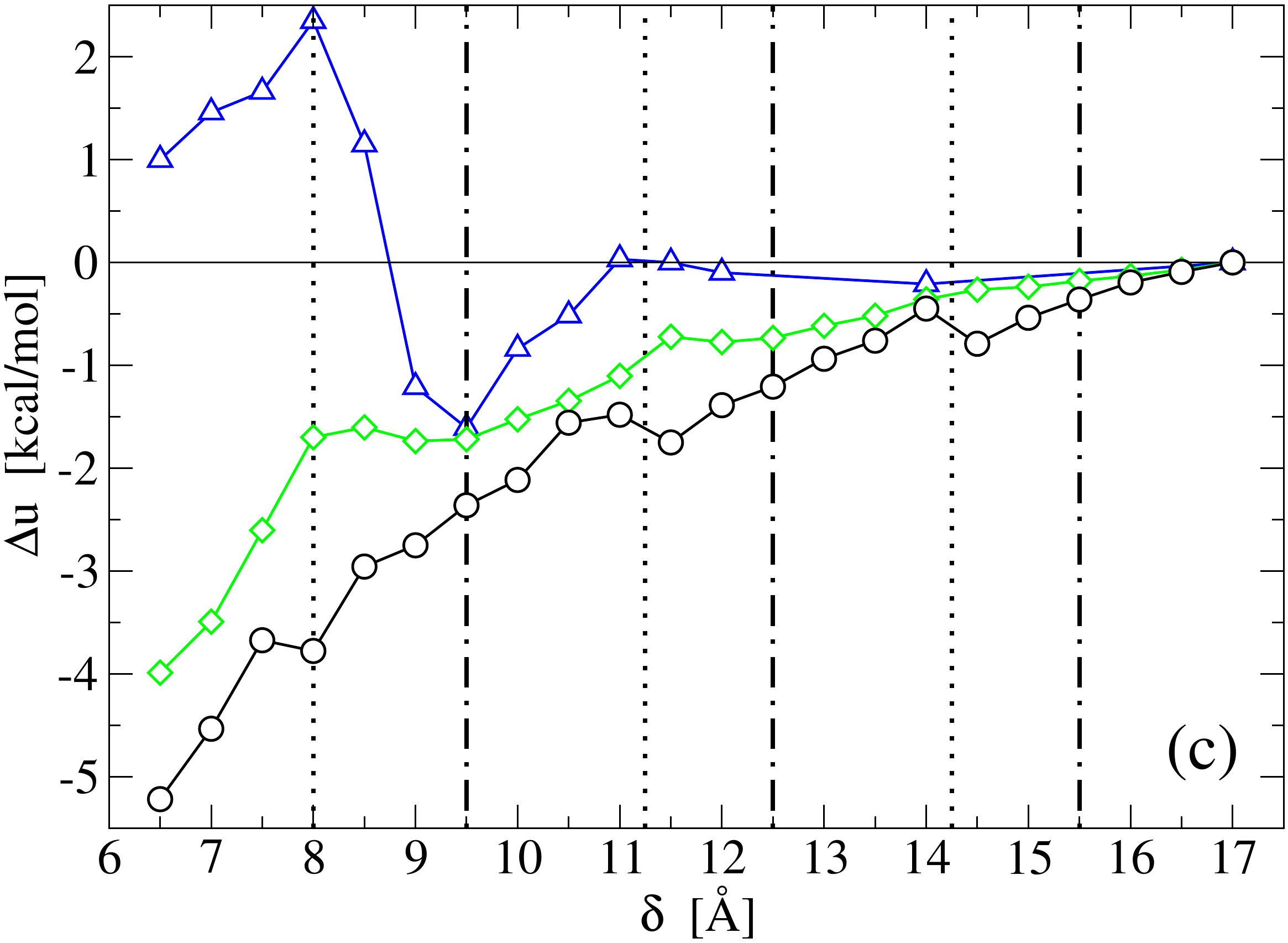}
\includegraphics[clip=true,width=7.5cm]{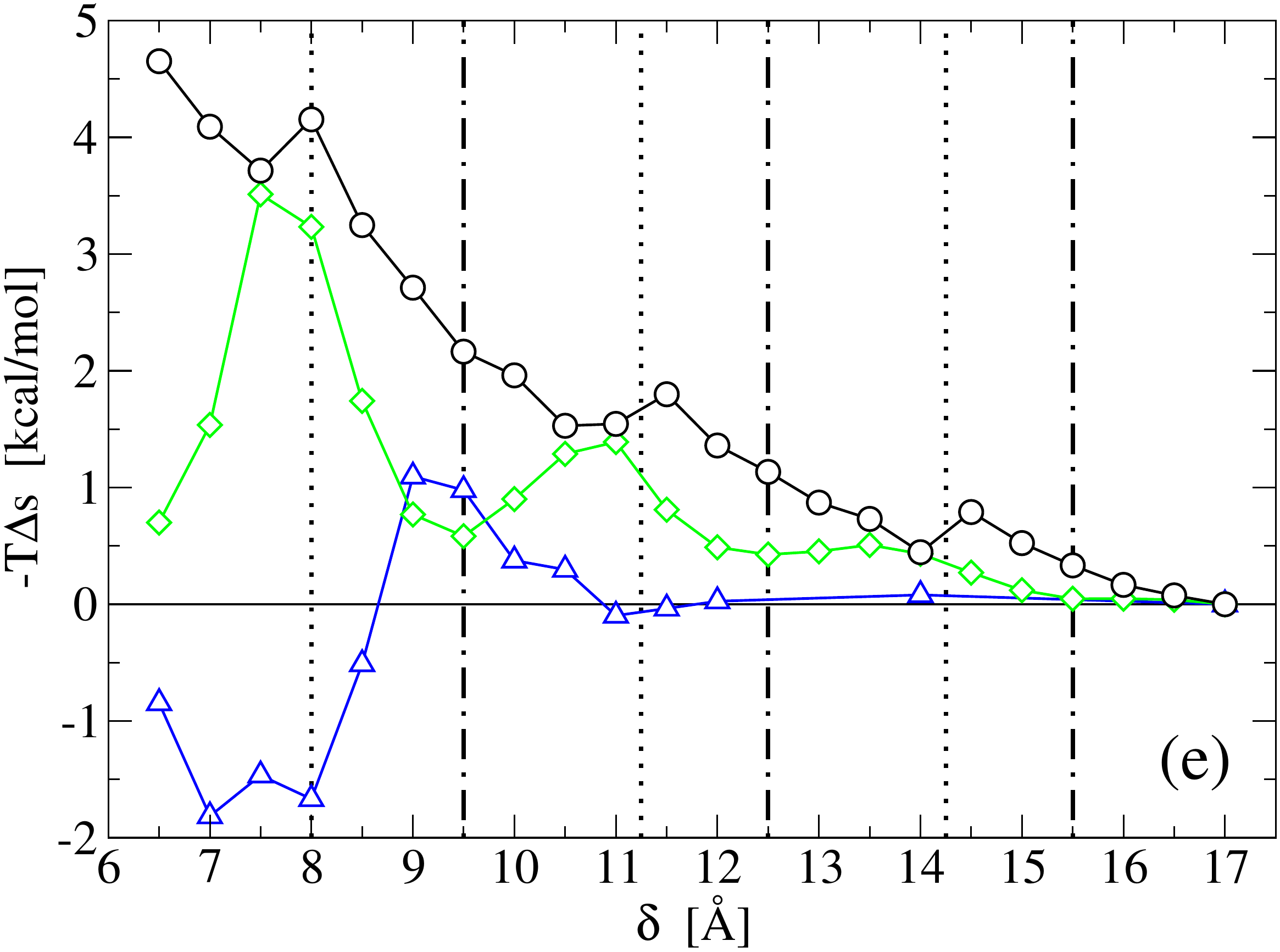}
\end{minipage}\hspace{0.2cm}
\begin{minipage}{7.5cm}
\includegraphics[clip=true,width=7.5cm]{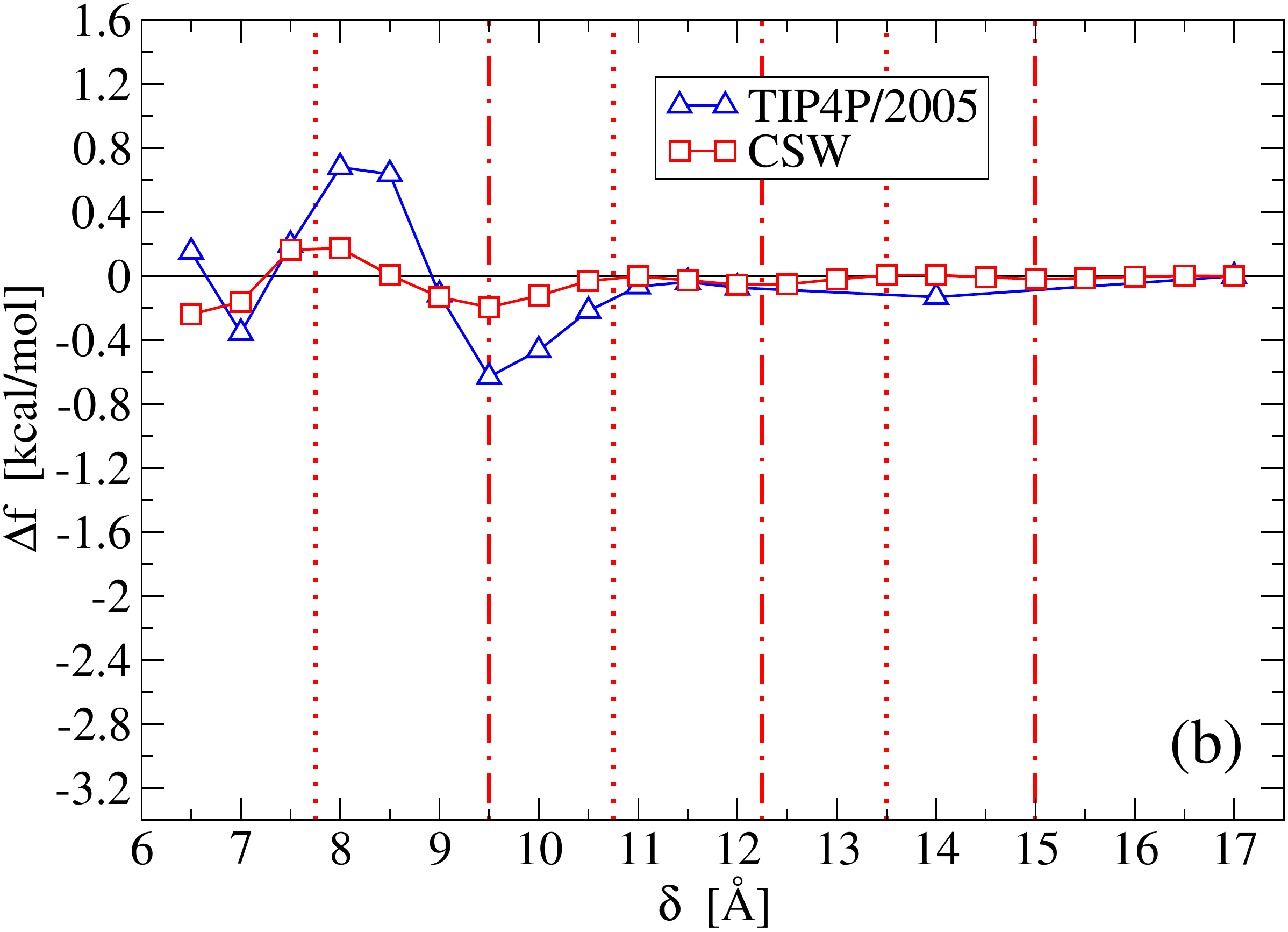}
\includegraphics[clip=true,width=7.5cm]{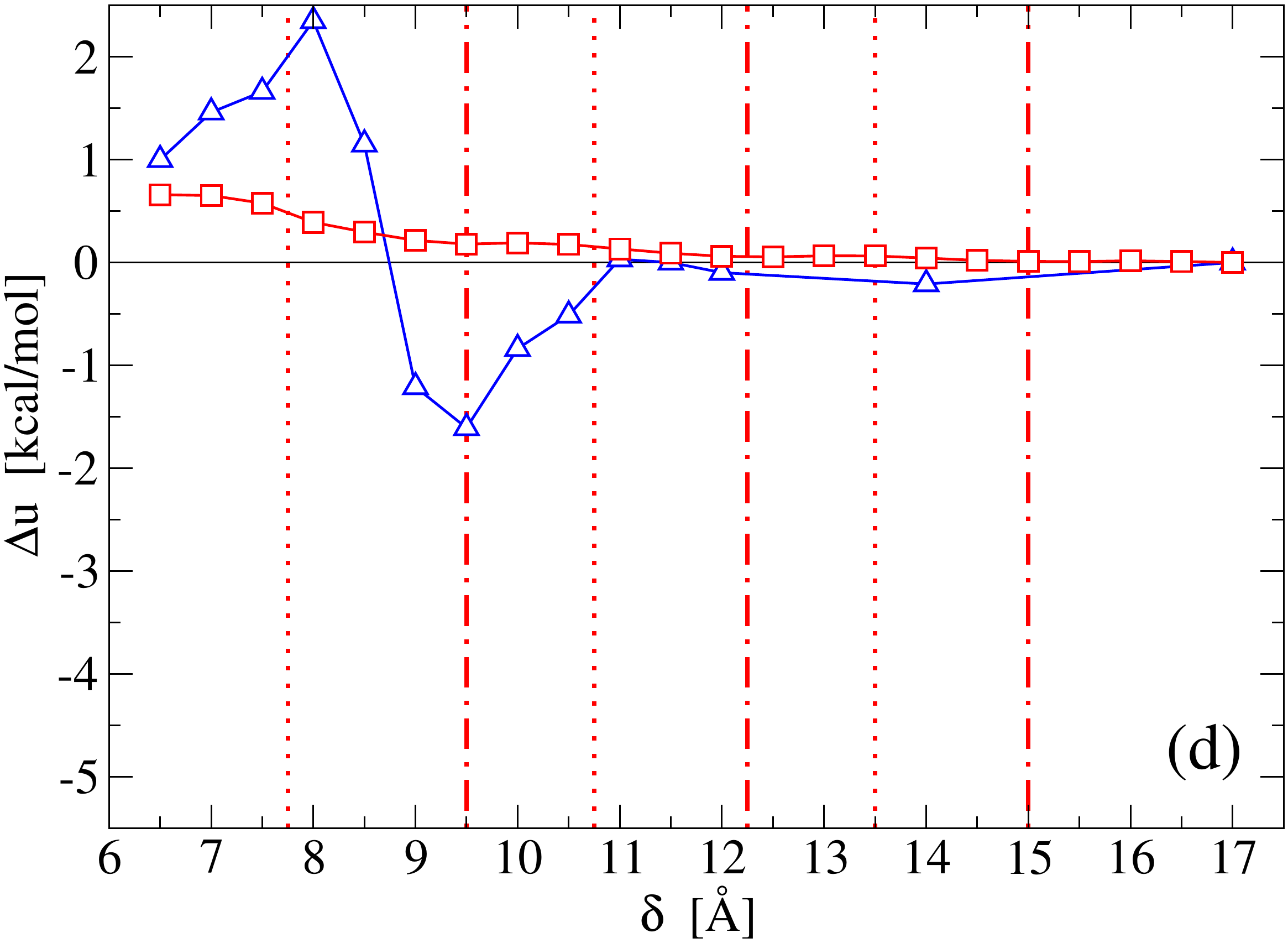}
\includegraphics[clip=true,width=7.5cm]{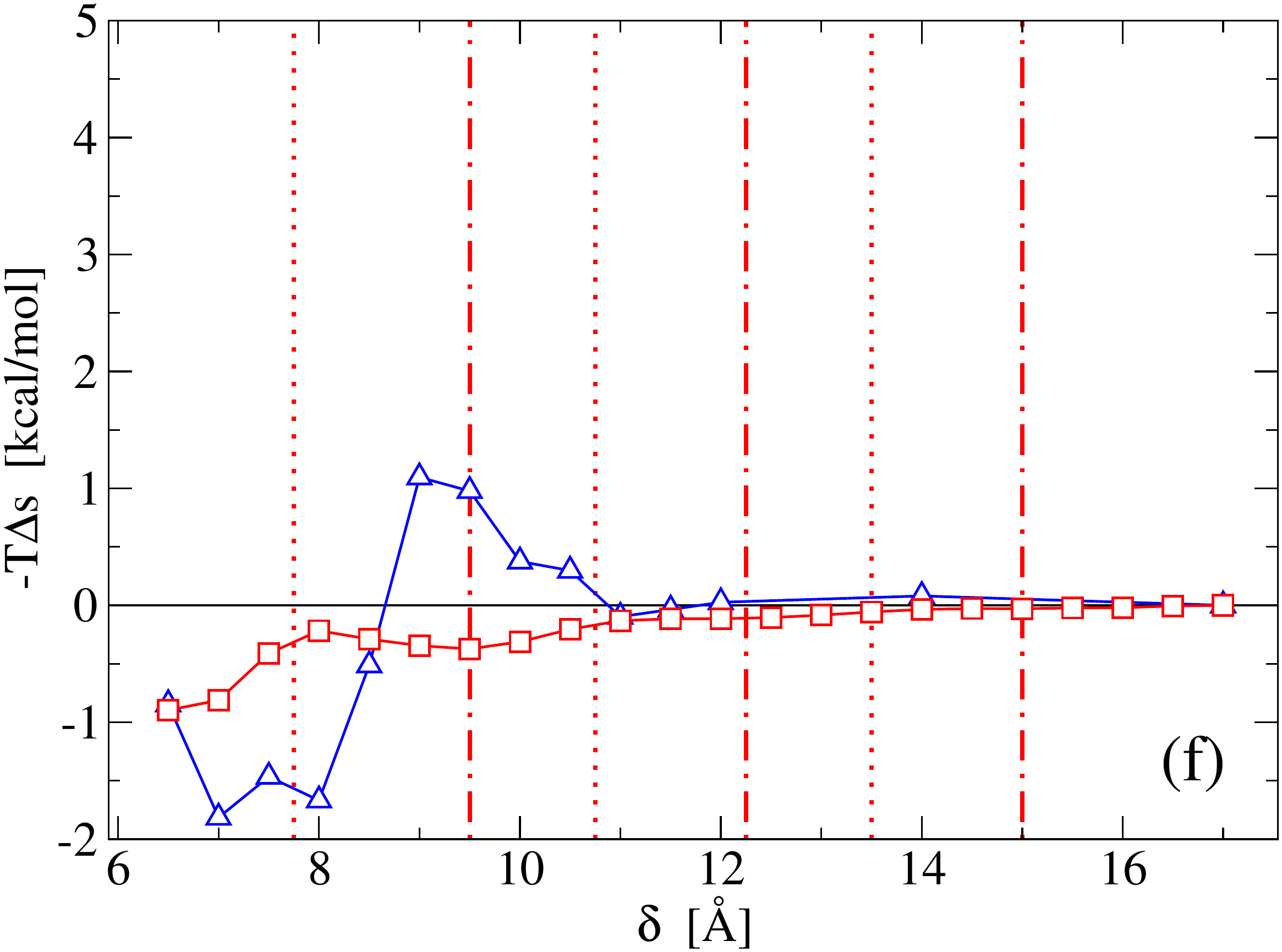}
\end{minipage}
\end{center}
\caption{Variation of (a, b) the free-energy density, $\Delta f$, (c, d) the internal energy density, $\Delta u$, and (e, f) the entropy density, $-T\Delta s$, for the confined  fluids when  the plate separation changes from $\delta_0=17$~{\AA} to  $\delta$.
Comparison of the TIP4P/2005-water (blue triangles) \cite{calero2020} with (left panels) the LJ with fluid-wall interaction energy 
$\epsilon_{w_1}=0.10$ kcal/mol (black circles) and 
$\epsilon_{w_2}=0.48$ kcal/mol (green diamonds), and 
(right panels) the CSW (red squares). In all the panels, vertical lines are as in Fig.\ref{fig:diff}, marking 
maxima (dotted lines) and minima (dot-dashed lines) in $D_\|$. We find that the lines of  $D_\|$-maxima and minima coincide with the $\Delta f$-maxima and minima, respectively, for the LJ   (left panels) and the CSW fluids (right panels).}
%Normalized free energy difference $\Delta f$, (a), (b), normalized internal energy difference $\Delta u$, (c), (d), and normalized entropy difference contribution $-T\Delta s$ to the free energy, (e), (f), as a function of the plate separation $\delta$ for the LJ and CSW fluids, respectively. In all panels we show the respective quantity for the TIP4P/2005 water model as reference.
%The fluid parameters are: $\rho_{LJ}=0.023~\AA^{-3}$ and $T_{LJ}=100K$ for the LJ fluid; $\rho_{CSW}=0.036~\AA^{-3}$ ($\rho^*_{CSW}=0.2$) and $T_{CSW}=100K$ ($T_{CSW}^*=1.0$) for the CSW fluid; $\rho_{TIP4P/2005}=0.033$~{\AA}$^{-3}$ and $T=300K$ for the TIP4P/2005.} 
\label{fig:energy1}
\end{figure*} 
%%%%%%%%%%%%%%%%%%%%%%%%%%

\subsubsection{Free Energy} 
Following Refs.~\cite{gao1997,gao1997b,calero2020}, 
we compute the macroscopic free-energy variation per particle, $ \Delta f$, 
as the macroscopic work done against the hydration forces   
to change the pore size from $\delta_0$ to $\delta$,
%by the hydration pressure
over the $N^s$ molecules, confined within the pore sub-volume of area $A^s$, 
as
 \begin{equation}\label{equ:F}
\Delta f(\delta)\equiv -A^s\int_{\delta_0}^{\delta}P_{\rm hydr}(\delta')/N^s(\delta')d\delta'.
\end{equation}
We numerically calculate $\Delta f(\delta)$ from the largest plates separation $\delta_0=17$~{\AA} to a generic value $\delta$, by setting in Eq.(\ref{equ:F})
$d\delta=0.5$~{\AA} as our minimal incremental value of $\delta$ (Fig.\ref{fig:energy1}.a, b).

Furthermore, we calculate the variation of the internal-energy per particle of the confined fluid, 
$\Delta u(\delta)\equiv U(\delta)/N^s(\delta)-U(\delta_0)/N^s(\delta_0)$, where $U(\delta)$ is the internal energy of the confined fluid at plates separation $\delta$ (Fig.\ref{fig:energy1}.c, d).
Finally, we estimate the variation of the entropy per particle of the confined fluid as 
$-T\Delta s(\delta)=\Delta f(\delta) - \Delta u(\delta)$ (Fig.\ref{fig:energy1}.e, f).

We find that the LJ and the CSW fluid present oscillations of $\Delta f$ in phase with those for the TIP4P/2005 water  \footnote{Apart from the oscillation around $\delta\simeq14$~{\AA} that for water is not observed, possibly, for lack of resolution.}. Furthermore, the CSW liquid and the LJ with weaker fluid-wall interaction (LJ$_w$, with $\epsilon_{w_1}=0.10$ kcal/mol)  are qualitatively very similar, with smoother oscillations for the CSW due to its pronounced soft core, as already observed for $D_{\|}$ (Fig.\ref{fig:diff}). 
Nevertheless, we observe important differences between the two isotropic fluids and the water.

First, the internal energy, $\Delta u$, and entropy, $-T\Delta s$ for the isotropic fluids oscillate but never change sign, while they do  for water. In particular, for the LJ$_w$ the $\Delta u$ is always negative and the $-T\Delta s$ is always positive, while for the CSW the signs are inverted. Nevertheless, the two contributions sum up in a similar $\Delta f$ for both isotropic fluids.

Second, for small pores  the $\Delta f$ for the isotropic fluids has deeper minima. Hence, their stability   increases for smaller pore sizes and is maximum for the monolayer. Instead, for water the deeper minimum of $\Delta f$ is for the bilayer \cite{calero2020}.

Third, for the LJ$_w$  the entropy variation  $-T\Delta s(\delta)$ is  positive and in average increases for decreasing $\delta$. 
Hence, the structural order of these confined liquids increases when the pore size decreases, 
consistent with our calculations of the longitudinal translational order $t_\|$ (Fifg.~\ref{fig:torderp}a).
For the CSW, $-T\Delta s(\delta)$ is negative and $t_\|$ is almost constant (Fifg.~\ref{fig:torderp}b), suggesting that the translational order  has a minor effect in the calculations of $-T\Delta s(\delta)$  for  the confined CSW.
For water, instead, $-T\Delta s(\delta)$ is negative 
for $\delta\lesssim 8.7$\AA\, for a confined monolayer,  
and positive for a confined bilayer, around  $\delta=9.5$\AA\  \cite{calero2020}. 
Hence, a confined water monolayer is less ordered than bulk water, while a water bilayer maximizes the  structural order.

Hence, comparing the three models, we can state that the more stable free energy minimum for water is the bilayer and it is energy driven and more structured than bulk. The monolayer of water is less stable and it is entropy driven.  
For the isotropic fluids the more stable free energy minimum is for the monolayer: for the  LJ$_w$  it is energy driven, while for the CSW it is entropy driven.

\subsubsection{Dependence of the Free Energy on Fluid-Wall Interaction}
Qualitative differences in the excess free energy between (SPC/E) water and a LJ fluid have been found also with density functional theory as a function of the fluid-wall interaction, although between face centered cubic (FCC)-structured slabs
\cite{lam2018}.  
Hence, to understand how our results depend on the fluid-wall interaction, we consider a LJ liquid with a strong wall-attraction energy  (LJ$_s$), with $\epsilon_{w_2}=0.48$ kcal/mol (green diamonds in left panels of Fig.~\ref{fig:energy1}). 

We find that the free energy oscillation for the LJ$_s$ are stronger than for the LJ$_w$, but the minima and maxima occurs, approximately, at the same pore sizes $\delta$. In particular, the entropy oscillations of the  LJ$_s$ are large, showing that the stronger fluid-wall interaction has a larger structural effect with respect to the   LJ$_w$ case.

\begin{figure*}%[t!]
\includegraphics[clip=true,width=7.5cm]{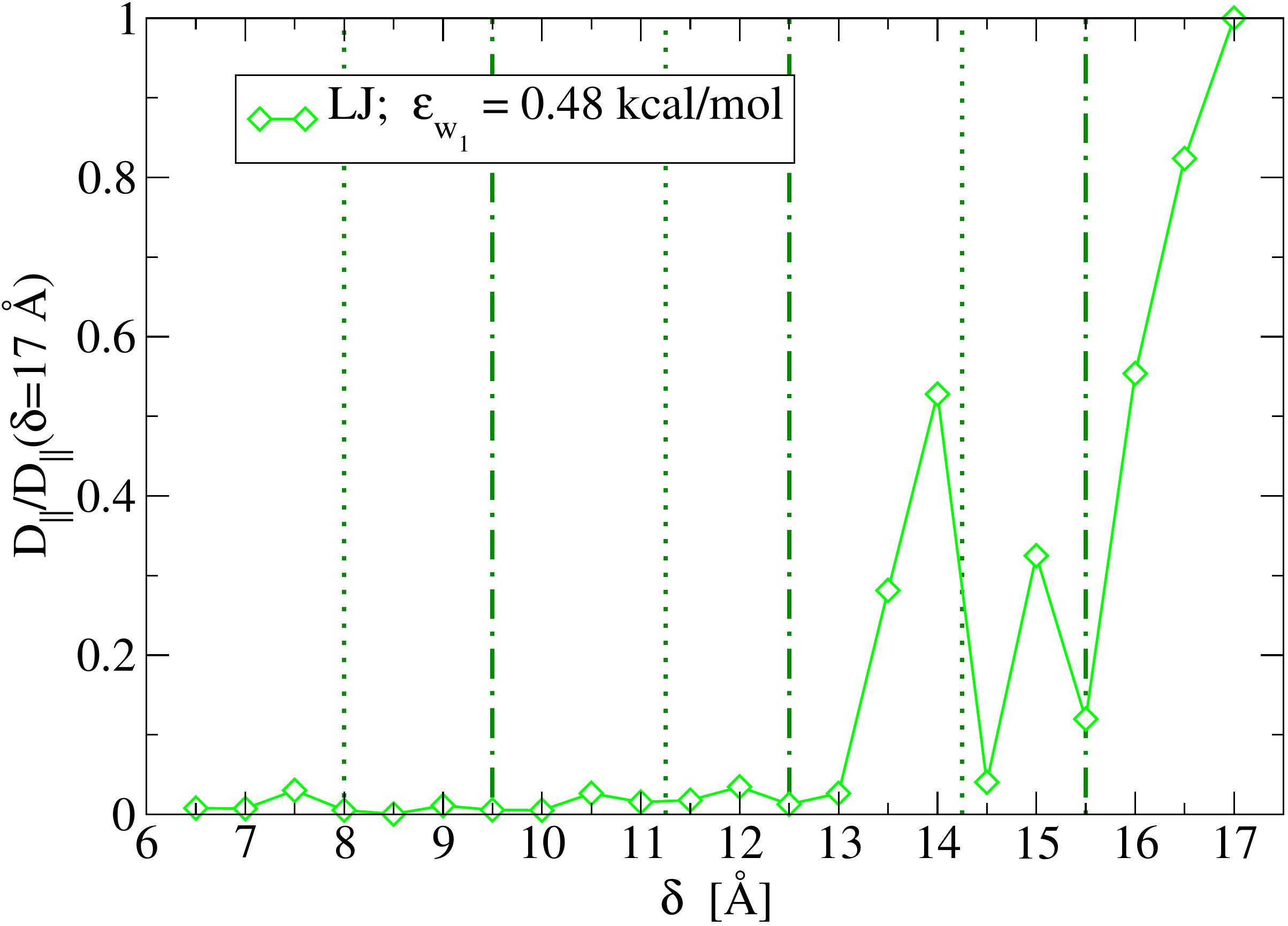}
\caption{Longitudinal diffusion coefficient $D_{\|}$, normalized to its large $\delta$ value, for the LJ$_s$ with strong fluid-wall interaction $\epsilon_{w_2}=0.48$ kcal/mol in a slit-pore,  as a function of the plate separation $\delta$. The vertical lines mark, approximately, maxima (dotted lines) and minima (dot-dashed lines) of $\Delta f$ for the LJ$_s$  fluid (Fig. \ref{fig:energy1}). The value of $D_{\|}$
at $\delta=17$~\AA\ is $\simeq 23$ nm$^2$/ns.}
\label{fig:S1}
\end{figure*}

This is confirmed when we calculate the longitudinal diffusion coefficient $D_{\|}$ for  the LJ$_s$ (Fig. \ref{fig:S1}). We find that, at variance with the  LJ$_w$ case (Fig. \ref{fig:diff}a), the LJ$_s$ freezes for $\delta \leq 13$~\AA. 
The parallel diffusivity inside the pore goes to zero when all the fluid layers are frozen, in a distorted triangular lattice, which happens when the peaks of the density profile are completely formed and there are no particles in between (Fig. S4 in Supplementary Material).

Crystallization and dynamic freezing have been found also for water confined into a graphene slit-pore 
when TIP4P/2005-water is at high pressure ($P=400$ bar) and a temperature ($T=275$ K) below the one considered here \cite{MCF2017}.  However, it occurs for a   bilayer that, as seen above, is the more stable configuration for confined water. Under these conditions, TIP4P/2005-water crystallizes into a hexagonal bilayer \cite{MCF2017} at a temperature that is much above the bulk melting temperature \cite{Conde:2013aa}\footnote{$T_{\rm m}(P=400 $bar$)< T_{\rm m}(P=1 $bar$)=249.5\pm 0.1 K$ \cite{Conde:2017aa}.}.  As a consequence of the large bilayer stability,  the confined crystal undergoes  reentrant melting  when the pore size allows only a water monolayer \cite{MCF2017}.

These results show that a strong fluid-wall interaction can induce crystallization in both confined LJ and water, however, they do not rationalize  the sub-nm speed-up and the bilayer strong-stability that we find in water. Hence, these properties are specific  of confined water, possibly related to its unique  hydrogen bonds. Indeed, the hydrogen bond network and its specific geometry are held responsible for the crystallization of sub-nm confined water into bilayer ices  at ambient conditions in experiments \cite{algara-siller2015, Zhou:2015tn, Algara-Siller:2015vv}
and simulations
\cite{Zangi2003a, Zangi2004, Han2010, Wang:2015tj, Zubeltzu:2016aa, PhysRevLett.116.085901, Zubeltzu:2017aa}
and its reentrant melting by changing the slit-pore size \cite{Zangi2003a, Zangi2004, MCF2017}.
To understand better how  it relates to the sub-nm speed-up and the bilayer strong-stability, we analyze the water  hydrogen bond network in the detail in the next section.

%the entropy of water is strongly dependent on the
%tetrahedrality contribution to the inter-particle interaction
%\cite{russo2018}, which is present in the TIP4P/2005 model unlike the  LJ and CSW fluids.   

\subsubsection{The Confined Water Hydrogen Bond Network}

\begin{figure*}%[t!]
\includegraphics[clip=true,width=8.5cm]{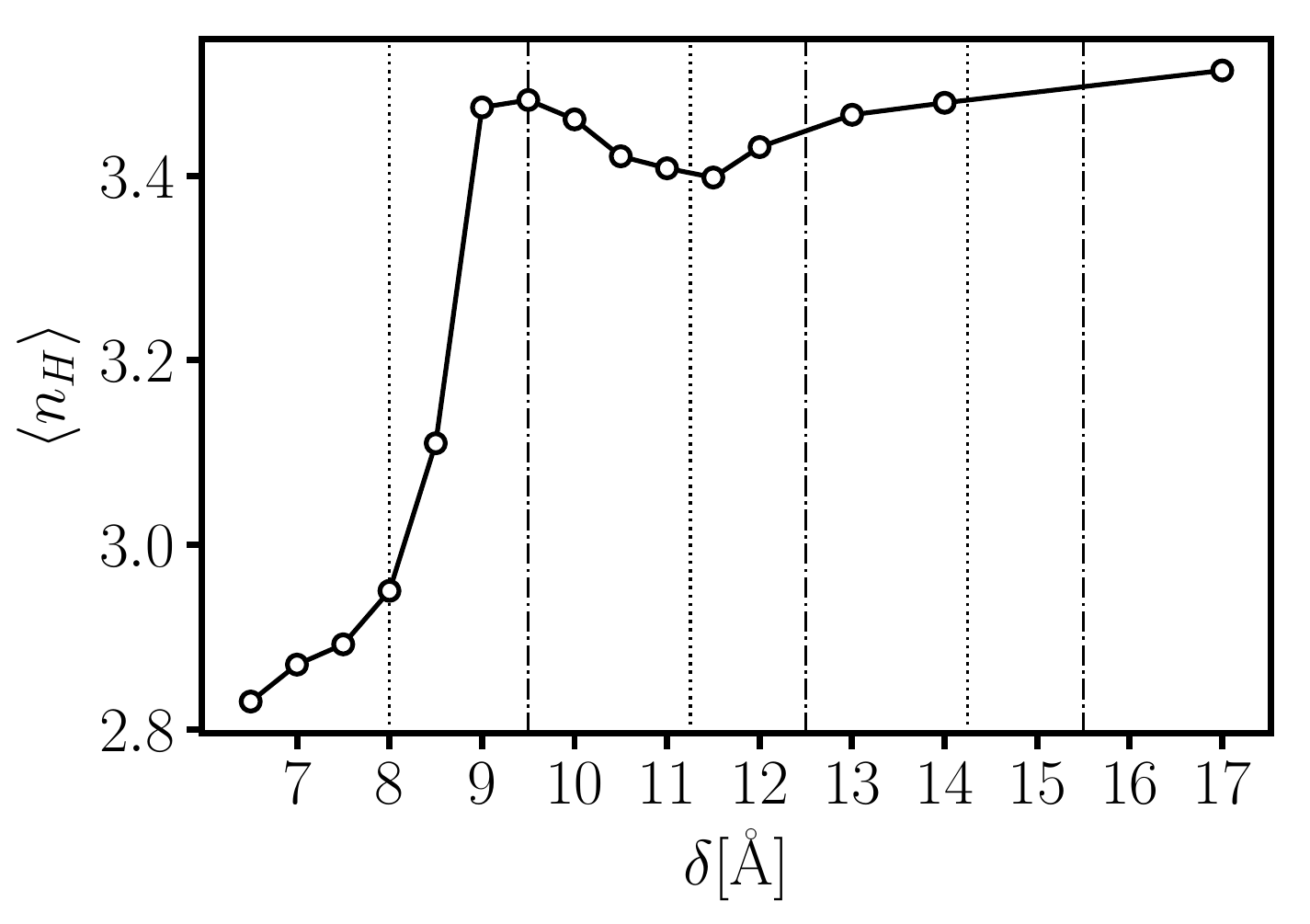}
\caption{The average number of hydrogen bonds per molecule, $\langle n_{\rm HB}\rangle$, for the water in the  confined subvolume, $V^s$, as a function of the pore size $\delta$. Vertical lines are defined as in Fig.~\ref{fig:diff}, approximately marking maxima (dotted lines) and minima (dot-dashed lines) in $D_\|$ and coinciding with $\Delta f$-maxima and minima, respectively, for the water in Fig.~\ref{fig:energy1} (left panels).}
\label{fig:nHB}
\end{figure*}

First we calculate the average number of hydrogen bonds per molecule, $\langle n_{\rm HB}\rangle$, for the water in the  confined subvolume, $V^s$, as a function of the pore size $\delta$ (Fig.~\ref{fig:nHB})\footnote{Vertical lines in the figure are defined for the LJ oscillations but, as discussed in the text, they approximate well the water oscillations.}. We find that $\langle n_{\rm HB}\rangle$ is almost as large as in bulk for the bilayer, where the free energy and $D_\|$ have their absolute minima. For other values of $\delta$,  $\langle n_{\rm HB}\rangle$ is smaller, with a local minimum at $\delta\simeq 11.5$~\AA, where both $\Delta f$ and $D_\|$  have local maxima.

These observation suggest that both diffusion and free energy are dominated, in these range of $\delta$, by the average number of hydrogen bonds. However, for  $\delta<9$~\AA\ the analysis is less intuitive. Indeed, The maximum in $D_\|$ at $\delta\simeq 8$~\AA\ does not correspond to a minimum in $\langle n_{\rm HB}\rangle$  (Fig.~\ref{fig:nHB}). Counterintuitively, for $\delta<8$~\AA, both $\langle n_{\rm HB}\rangle$ and $D_\|$ decrease.  

\begin{figure*}%[t!]
\includegraphics[clip=true,width=12cm]{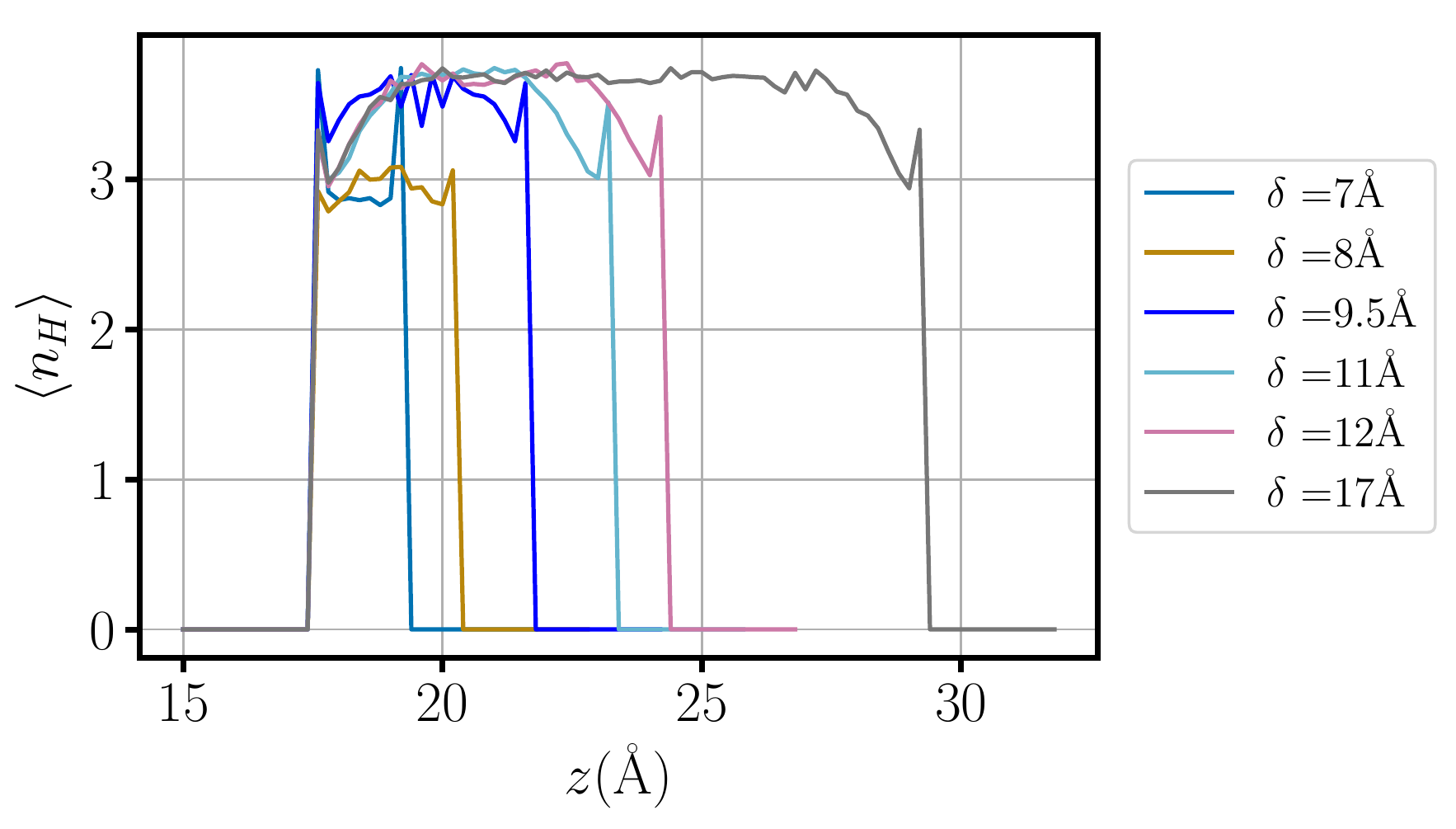}
\caption{The hydrogen-bond profile as a function of the water molecule position along the direction $z$ perpendicular to the slit-pore walls, for $7\leq \delta/$\AA$\leq 17$. 
For $\delta\geq 9.5$~\AA\ the profile saturates in its center to the bulk value $\langle n_{\rm HB}\rangle\simeq 3.5$, while it is less near the walls.  For $\delta\leq 8$~\AA, it is  $\langle n_{\rm HB}\rangle \simeq 3$ in the center and $\langle n_{\rm HB}\rangle \simeq 3.5$ near the walls.
Colors for each $\delta$ are indicated in the legend. For sake of comparison, for each $\delta$ the first peak of the profile is shifted at  $z_0=17.5$~\AA. The distance from the first peak and the nearest wall can be estimated from Fig.5 of Ref.~\cite{calero2020}--very similar to Figs. S1 and S2 for the LJ--and changes with $\delta$. The thickness of the hydrogen-bond profile changes with the thickness of the density profile in the same figures.}
\label{fig:profileHB}
\end{figure*}

This is the range of $\delta$-values where the confined-water free-energy is dominated by its entropy. In particular, its $-T\Delta s$ has a (structured) minimum for $7\lesssim \delta/$\AA$\lesssim 8$ (Fig.~\ref{fig:energy1}.e). Although not evident from the averaged $\langle n_{\rm HB}\rangle$, our detailed analysis  shows that, for these values of $\delta$,  the hydrogen-bond profile  is quite different from the cases at larger $\delta$. We find (Fig.~\ref{fig:profileHB}) that the hydrogen-bond profile for 
$\delta> 8$~\AA\ (with two or more layers) saturates in its center to a bulk-like value within $\simeq 4.5$~\AA\ from the graphene wall. For $\delta\leq 8$~\AA\  there is not enough space in the pore to allow the water molecules to arrange in such a saturated network. As a consequence, away from the wall, the profile reaches a local value of $\langle n_{\rm HB}\rangle \simeq 3$, indicating a less connected network. 

\begin{figure*}%[t!]
\includegraphics[clip=true,width=14cm]{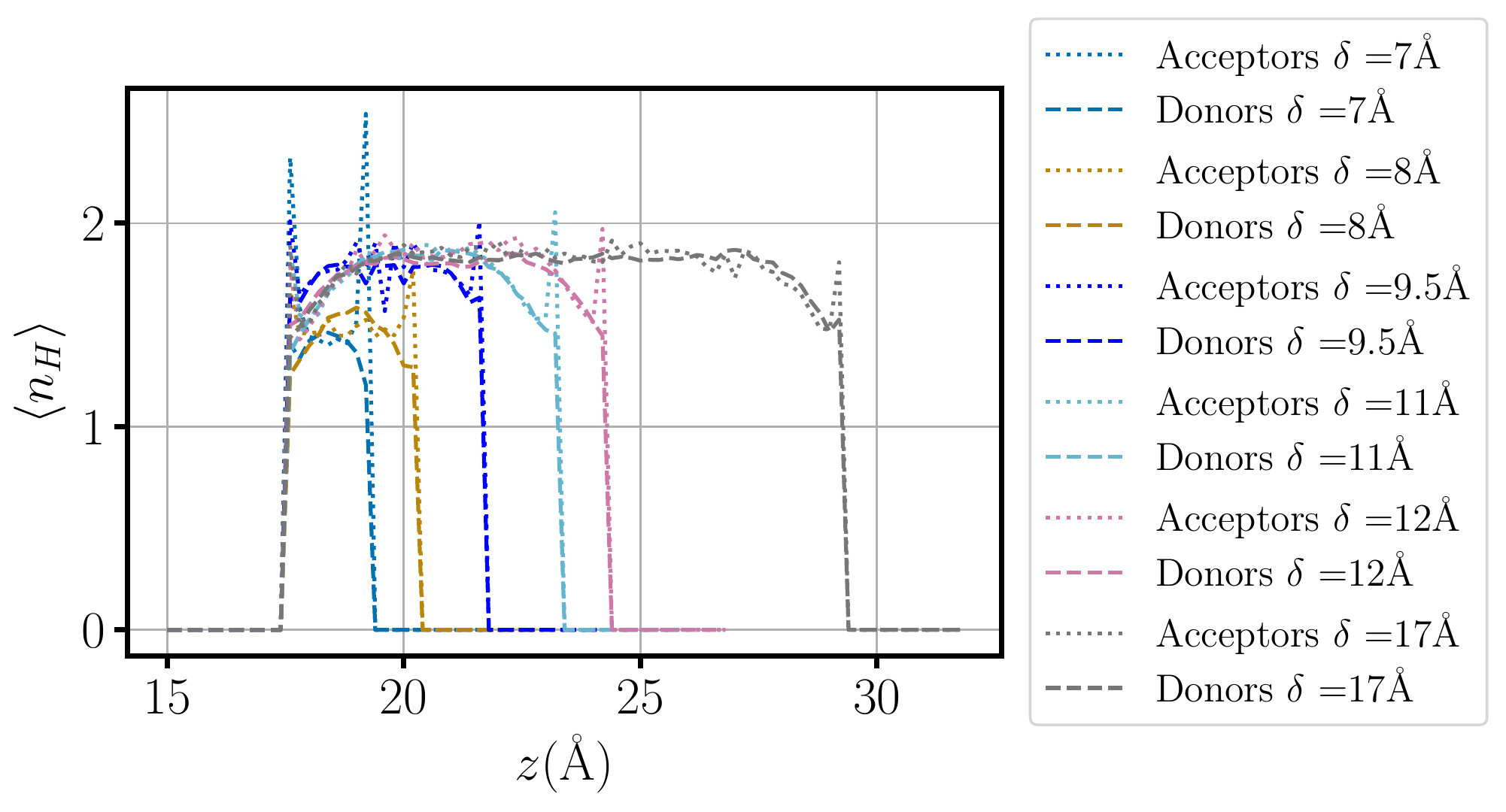}
\caption{The profiles of hydrogen-bond  acceptors (dotted lines) and donors  (dashed lines) as a function of the water molecule position along the direction $z$ perpendicular to the slit-pore walls, for $7\leq \delta/$\AA$\leq 17$. Colors are as in Fig.~\ref{fig:profileHB}.}
\label{fig:acc_don}
\end{figure*}

In particular, we calculate the profiles of donors and acceptors for the hydrogen bonds as a function of $z$, for each $\delta$ (Fig.~\ref{fig:acc_don}). We find that for $\delta=7 $~\AA\ (monolayer) the majority of the water molecules have their hydrogens pointing toward the center of the pore, away from the hydrophobic walls, as one would expect. This asymmetry between the donors and acceptors profiles smoothen for $\delta>8 $~\AA. The strong asymmetry for $\delta\leq 8$~\AA\ indicates that the hydrogen-bond network is hindered by the hydrophobic wall, facilitating the breaking of the Cooperative Rearranging Regions and the diffusion in confined water~\cite{delosSantos2012}. This observation is consistent with the larger entropy of the confined water monolayer with respect to the cases with more, well formed layers (Fig.~\ref{fig:energy1}.e).

%%%%%%%%%%%%%%%%%%%%%%%%%%%%%%%%%%%%%%%%%%%%%%%%%%%%%%%%%%%
%
%     SUMMARY AND DISCUSSION
%
%%%%%%%%%%%%%%%%%%%%%%%%%%%%%%%%%%%%%%%%%%%%%%%%%%%%%%%%%%%

\section{Conclusions}
\label{sec:conclusions}

We compare structure, dynamics and thermodynamics of  water confined in a graphene slit-pore 
with two isotropic liquids, a simple liquid (LJ) and an anomalous liquid (CSW), under similar conditions. We find that  below $\simeq 1$~nm, where only two or one layer can be accommodated,   
confined water is unique for, at least, the following reasons. 
\begin{itemize}
\item[(i)] Water goes from very large  to very small order changing the pore size from 0.95~nm to 0.80~nm, when compared with the bulk.
The considered isotropic liquids, instead, have a structural order that, although oscillating, increases in its maxima for decreasing pore size.
\item[(ii)] Water goes from less to more diffusive than bulk changing the pore size from 0.95~nm to 0.80~nm, with a maximum at $0.8$~nm. 
The isotropic liquids, instead, have a thermal diffusion  oscillating with the pore size, but with an overall decreasing diffusion coefficient for decreasing pore size.
\item[(iii)] Water has its maximum stability for the double layer at 0.95~nm, where it saturates its hydrogen-bond network. The monolayer at $\simeq 0.7$~nm is less stable and more disordered, with its hydrogen-bond network hindered by the hydrophobic graphene walls.
For the isotropic liquids, instead,  a  monolayer is more stable than two or more confined layers. While for the simple LJ the internal energy of the confined liquid is the leading contribution to the stability, for the anomalous liquid, CSW, it is the entropy, resembling more water. 
\end{itemize}

Our analysis  clarifies that these differences are all due to the water hydrogen-bond network. Therefore, the layering alone is not able to rationalize the properties of water under sub-nm confinement, not even if 
a stronger interaction with the walls is considered. We find that strong LJ-wall interaction  leads to freezing and crystallization at sub-nm pore-size, with an effect similar to a decrease of temperature for confined water \cite{calero2020} and opposite to increase in diffusion or disorder. 
 
Nevertheless, it is intriguing to observe that the differences with isotropic liquids fade out for pore with more than two layers (> 1nm). This is especially true for the CSW anomalous liquid, that in its bulk version, has some water-like properties \cite{OFNB08,leoni2014,leoni2016}, although not the entropy balance observed in water \cite{vilaseca:084507, Vilaseca2011}. However, the differences are emphasized for monolayers and bilayers.
Because water monolayers and bilayers are common in biology and nanofluidics, our results  pave the way to better understand how biology takes advantage of the unique water properties and how nanotechnology could mimic, in this respect, Mother Nature.

%%%%%%%%%%%%%%%%%%%%%%%%%%%%%%%%%%%%%%%%%%%%%%%%%%%%%%%%%%%
%
%     Methods
%
%%%%%%%%%%%%%%%%%%%%%%%%%%%%%%%%%%%%%%%%%%%%%%%%%%%%%%%%%%%

\section{Methods}
\label{sec:methods}

\subsection{Confined LJ fluid}
We simulate particles interacting via a LJ potential (Fig.~\ref{fig:snapshot}.b)
\begin{equation}
U_{\rm LJ}(r)  \equiv
4\epsilon_{\rm LJ}\left[
\left(\frac{\sigma_{\rm LJ}}{r}
\right)^{12} 
-
\left(\frac{\sigma_{\rm LJ}}{r}
\right)^6
\right]
\label{LJ}
\end{equation}
with 
$\sigma_{\rm LJ}=3.16$~\AA\ and 
$\epsilon_{\rm LJ}=0.2$ kcal/mol.
These parameters are chosen in such a way to compare with the  
 LJ contribution of the TIP4P/2005 potential (with same size and 0.185 kcal/mol as LJ energy)
 \cite{Abascal:2005bh}.
%(c.f.r for argon is $\epsilon_{LL}\simeq 0.24$ kcal/mol and $\sigma_{LL}\simeq 3.33$~{\AA} \cite{Rowley:1975aa, White:1999aa}). 
In order to reduce the computational cost, we impose a cutoff for the interaction potential at a distance $r_c=10$~{\AA}.

The slit-pore is composed of two parallel graphene sheets. 
Each sheet is a honeycomb lattice made of 
$N_{\rm G}=960$ frozen particles, 
with inter-particle distance $1.42$~\AA,  
lateral sizes $L_x = 49$~\AA, $L_y=51$~\AA, 
and an area   $A\equiv L_x \times L_y\simeq  25$ nm$^2$.
The graphene particles of the walls interact with the fluid particles through a LJ potential as in 
Eq.(\ref{LJ}) \cite{charmm2010}
with size $\sigma_{w}=3.26$~\AA, 
and energy 
$\epsilon_{w_1}=0.1$ kcal/mol (case 1, weaker than the fluid-fluid interaction), 
or
$\epsilon_{w_2}=0.48$ kcal/mol (case 2, stronger than the fluid-fluid interaction).
The two choices, $\epsilon_{w_1}$ and $\epsilon_{w_2}$, allow us 
to study the effects of the fluid-wall interaction strength.
In particular, we chose $\epsilon_{w_2}=2.4 \times \epsilon_{LJ}$ to compare our results with those  by Gao et al.  \cite{gao1997,gao1997b}.

%For systems $i$ and $ii$ we use LAMMPS \cite{lammps} with $N_{\rm tot}=25000$ particles. 

We perform NPT simulations at constant number $N_{\rm tot}=25000$ of LJ particles, constant temperature $T=100$ K, and constant bulk pressure $P_{\rm bulk}=1$ atm, 
leaving the box volume, $V \equiv L_x^{\rm box} \times L_y^{\rm box} \times L_z^{\rm box}$ with $L_x^{\rm box}=L_y^{\rm box}$, free to change \footnote{At this state point, corresponding to a bulk number density $\rho_{\rm bulk}\simeq 0.023$~\AA$^{-3}$, i.e., a reduced density $\rho_{\rm bulk}^*\equiv \rho_{\rm bulk}\sigma_{LJ}^3\simeq 0.73$, the bulk is liquid and the confined region is filled with fluid.}.

%In the confined sub-volume $V^s$ the available space changes with $T$, as discussed in Ref. \cite{Engstler:2018ab} and references therein, hence, the properties  for the confined fluid are calculated at constant $T$, constant bulk pressure $P_{\rm bulk}$, and constant  chemical potential $\mu$.
%

%The simulation time step for both LJ and CSW corresponds to $10^{-15}$s. 
We simulate the system with LAMMPS, adopting the Nose-Hoover thermostat and barostat  \cite{lammps}, with relaxation time $10^2$ and $10^3$ MD steps, respectively, and with $10^5$ MD steps of relaxation, enough to reach equilibrium in the bulk and within the  confined sub-region. Next 
we compute the observables for $10^5$ more MD steps, recording each quantity every $10^3$ MD steps. 

\subsection{Confined CSW fluid}
We describe the anomalous fluid with the CSW potential  
(Fig.~\ref{fig:snapshot}.b)
\cite{Fr07a,OFNB08,vilaseca:084507,leoni2014,leoni2016}  
\begin{equation}
\begin{array}{lll}
U_{\rm CSW}(r) & \equiv & \dfrac{U_R}{1+\exp(\Delta(r-R_R)/a)}\\
&&\\
&& -U_A\exp\left[-\dfrac{(r-R_A)^2}
  {2\omega_A^2}\right]+U_A\left(\dfrac{a}{r}\right)^{24}  
\end{array}
\end{equation}
where $a$ is the diameter of the particles, 
$R_A$ and $R_R$ are the
distance of the attractive minimum and the repulsive radius,
respectively, 
$U_A$ and $U_R$ are the energies of the attractive well
and the repulsive shoulder, respectively, 
$\omega_A^2$ is the variance of the Gaussian centered in $R_A$ and 
$\Delta$ is the parameter which controls the slope between the shoulder and the well at $R_R$.
We choose the CSW parameters in such a way that the resulting potential  compares at the best with  LJ potential (Fig.~\ref{fig:snapshot}.b):
$U_A=0.2$ kcal/mol, 
$a=1.77$~\AA, 
$R_A=2a\simeq3.54$~\AA, 
$U_R/U_A=2$, 
$R_R/a=1.6$, 
$(\omega_A/a)^2=0.1$,
% as in Ref.~\cite{OFNB08,Fr07a,vilaseca:084507,leoni2014,leoni2016}, 
$\Delta=30$, 
%as for the bulk case studied in Ref.\cite{vilaseca:084507},
and a cutoff at a distance $r_c=10$~{\AA}.

We adopt the same slit-pore as for the LJ fluid with the weak fluid-wall interaction, $\epsilon_{w_1}=0.1$ kcal/mol,  simulating  the system with LAMMPS and Nose-Hoover thermostat \cite{lammps}, with the same equilibration and production statistics as for the LJ. We perform the simulations  at constant number $N_{\rm tot}=25000$ of CSW particles, constant temperature $T$, and constant box volume $V$, leaving the bulk pressure $P_{\rm bulk}$ free to change \footnote{Simulations at constant $P_{\rm bulk}$ for the CSW fluid at the same $T$,  same $P_{\rm bulk}$  and same $N_{\rm tot}$ as for the LJ fluid would require a much larger box for the CSW than the LJ  in order to get a comparable number of
particles inside the sub-volume $V^s$.}. We consider different values of temperature, $T/$K=60, 80, 100, and we 
vary $L_x^{\rm box} = L_y^{\rm box}$, changing the box section parallel to the slit-pore plates, to control the bulk number density as $\rho_{\rm bulk}/$\AA$^{-3} =$ 0.027, 0.036, 0.045, 0.054, i.e., reduced densities
$\rho_{\rm bulk}^*\equiv \rho_{\rm bulk}a^3 =$ 0.15, 0.2, 0.25, 0.30,
  all corresponding to the bulk liquid phase \cite{vilaseca:084507}
(Fig.~S1 ans S2 in Supplementary Material).
We focus on  the state point 
at $\rho_{\rm bulk}=0.036$~{\AA}$^{-3}$ and $T=100$ K because it shows a  dynamics 
comparable to the confined LJ, as discussed in the main text.

\subsection{Confined TIP4P/2005 water}
For the confined TIP4P/2005 water \cite{Abascal:2005bh}, we use the data and the parameters as described in Ref.\cite{calero2020}. 
Specifically,  the system has $N_{\rm tot}=2796$ water molecules in a   box  with constant volume $V=4.2\times 4.2\times 5.1$ nm$^3$ and constant  $T=300$ K, corresponding  to $P_{\rm bulk}=400$ atm and a density $\rho_{\rm bulk}\approx 1$ g/cm$^3$, i.e., a number density $0.033$~{\AA}$^{-3}$. 
The graphene slit-pore has two $24.6$~\AA\ $ \times 25.5$~\AA\ rigid plates, and water-carbon interactions modeled as LJ potential as in  the  CHARMM27 force field,  adopting the Lorentz-Berthelot rules,  a cut off of the Van der Waals interactions at 12~\AA, a smooth switching function
starting at 10~\AA, and the particle mesh Ewald method \cite{Essmann_JCP1995}, with a grid space of $\approx 1$~\AA, for the calculation of the long-range electrostatic forces.
%The system is at constant $N$, $V$, and $T$, equilibrating for 5 ns with a Berendsen thermostat~\cite{Berendsen_JPhysChem1984}. Then, we collect data every 10 ps for the next 50 ns and every 0.1 ps for the next  8 ns with GROMACS~\cite{Gromacs4}. We compute  using 
%We analyze water molecules within the central sub-region between the plates 
As for the fluids (A) and (B), also  in this case, the observables  are calculated in 
a confined sub-volume $V^s$, at constant $T$,  and constant  chemical potential $\mu$. Further details are given in Ref.\cite{calero2020}.

\begin{acknowledgement}
We acknowledge support from the Spanish grant PGC2018-099277-B-C22 
(MCIU/AEI/ERDF). G.F. acknowledges support from the ICREA Foundation (ICREA Academia prize). 
\end{acknowledgement}

%\bibliographystyle{prsty}
%\bibliography{biblioteca-ll-isotropic_31-01-2018}% Produces the bibliography via BibTeX.
\bibliography{biblio}

\end{document}